\documentclass[conference]{IEEEtran}
 
\usepackage{etex}
\usepackage[vlined,ruled]{algorithm2e}
\usepackage[dvipsnames]{xcolor}
\usepackage{pgfpages}
\usepackage[cmex10]{amsmath}		%Load Fonts
\usepackage{amssymb, mathrsfs}
\usepackage{cite}
\usepackage{url}

\usepackage{dsfont}
\usepackage{siunitx}
\usepackage{verbatim}									%verbatim and comment environemtns		
\usepackage{mathtools}									%used for xmapsto command
\usepackage{siunitx}									%load SI units
\usepackage{mdframed}

\usepackage{graphicx}
\usepackage{epstopdf}

\usepackage[font=small]{caption}
\usepackage{subcaption}
\usepackage[activate={true,nocompatibility},final,tracking=true,kerning=true,spacing=true,factor=1100,stretch=10,shrink=10]{microtype}

\usepackage{array}
\usepackage{multirow}
\usepackage{enumerate}
\usepackage{booktabs}

\graphicspath{{./fig/}}

\usepackage{pgfplots}
\pgfplotsset{width=8cm,compat=newest}
\newcommand*{\damping}{0.006}%
\newcommand*{\freq}{25}%
\pgfmathsetmacro{\freqd}{sqrt(1-(\damping)^2)*\freq}%

\pgfplotsset{
    standard/.style={
    axis x line=middle,
    axis y line=middle,
    enlarge x limits=0.15,
	enlarge y limits=0.15,
	every axis plot post/.style={mark options={fill=black}},
	}
}

\pgfplotsset{%
    ,compat=1.12
    ,every axis x label/.style={at={(current axis.right of origin)},anchor=north west}
    ,every axis y label/.style={at={(current axis.above origin)},anchor=north east}
    }

\usepackage{tikz}
\usetikzlibrary{arrows}
\usetikzlibrary{decorations.pathmorphing}
\usetikzlibrary{decorations.markings}
\usetikzlibrary{snakes}
\usetikzlibrary{matrix}
\usetikzlibrary{calc}
\usetikzlibrary{patterns}
\usetikzlibrary{positioning}
\usetikzlibrary{shapes}
\usetikzlibrary{shapes.symbols}
\usetikzlibrary{shapes.geometric}
\usetikzlibrary{shadows.blur}
\usetikzlibrary{shapes.arrows}
\usetikzlibrary{shapes.multipart}
\usetikzlibrary{shapes.callouts}
\usetikzlibrary{shapes.misc}

\tikzstyle{every node}=[font=\small]
\tikzstyle{every path}=[line width=0.8pt,line cap=round,line join=round]

\usepackage[american,smartlabels]{circuitikz}
\ctikzset{bipoles/thickness=1}
\ctikzset{bipoles/length=0.8cm}
\ctikzset{bipoles/diode/height=.375}
\ctikzset{bipoles/diode/width=.3}
\ctikzset{tripoles/thyristor/height=.8}
\ctikzset{tripoles/thyristor/width=1}
\ctikzset{bipoles/vsourceam/height/.initial=.7}
\ctikzset{bipoles/vsourceam/width/.initial=.7}

\newcommand{\real}{\mathbb{R}}

\newcommand{\integer}{\mathbb{Z}}

\newcommand{\vect}[1]{\mathbbold{#1}}

\newcommand{\vzeros}[1][]{\vect{0}_{#1}}
\DeclareSymbolFont{bbold}{U}{bbold}{m}{n}
\DeclareSymbolFontAlphabet{\mathbbold}{bbold}

\renewcommand{\theenumi}{(\roman{enumi})}

\newcommand\oprocendsymbol{\hbox{$\square$}}
\newcommand\oprocend{\relax\ifmmode\else\unskip\hfill\fi\oprocendsymbol}

\newtheorem{theorem}{Theorem}[section]
\newtheorem{lemma}[theorem]{Lemma}

\newtheorem{proposition}[theorem]{Proposition}

\newcommand{\define}{\coloneqq}

\newif\ifforstudents

\ifCLASSINFOpdf
\else
\fi
\usepackage{url}
\usepackage{xcolor, soul}
\sethlcolor{green}

\usepackage[]{fancyhdr} % 
\newcommand{\changefont}{\fontsize{9}{9}\selectfont}
\usepackage{tikz}
\usetikzlibrary{spy}
\usetikzlibrary{backgrounds}
\usetikzlibrary{arrows}
\usetikzlibrary{arrows.meta}
\usetikzlibrary{decorations.pathmorphing}
\usetikzlibrary{decorations.markings}
\usetikzlibrary{snakes}
\usetikzlibrary{matrix}
\usetikzlibrary{calc}
\usetikzlibrary{patterns}
\usetikzlibrary{positioning}
\usetikzlibrary{shapes}
\usetikzlibrary{shapes.symbols}
\usetikzlibrary{shapes.geometric}
\usetikzlibrary{shadows.blur}
\usetikzlibrary{shapes.arrows}
\usetikzlibrary{shapes.multipart}
\usetikzlibrary{shapes.callouts}
\usetikzlibrary{shapes.misc}
\usetikzlibrary{fit}
\usepackage{makecell}

\tikzstyle{every node}=[font=\small]
\tikzstyle{every path}=[line width=0.8pt,line cap=round,line join=round]

      \tikzset{
  basic box/.style = {
    shape = rectangle,
    align = center,
    draw  = #1,
    fill  = #1!5,
    rounded corners}}

\IEEEoverridecommandlockouts
\begin{document}

\title{Data-Driven Fast Frequency Control using Inverter-Based Resources }

\author{
\IEEEauthorblockN{E.~Ekomwenrenren}
\IEEEauthorblockA{Dept. Electrical \& Comp. Eng.\\University of Waterloo\\
Waterloo, ON, Canada}
\and
\IEEEauthorblockN{J.~W.~Simpson-Porco}
\IEEEauthorblockA{Dept. Electrical \& Comp. Eng.\\University of Toronto\\
Toronto, ON, Canada}
\and
\IEEEauthorblockN{E.~Farantatos, M.~Patel, A.~Haddadi, L. Zhu}
\IEEEauthorblockA{Grid Operations \& Planning Division\\Electric Power Research Institute\\
Palo Alto, CA, USA}
\thanks{This work was supported by EPRI Project \#10009168: Wide-Area Hierarchical Frequency and Voltage Control for Next Generation Transmission Grids, and by NSERC Discovery Grant RGPIN-2017-04008.}}


%Telephone: (800) 555--1212, Fax: (888) 555--1212}

% <-this % stops a space


%\lhead{11TH BULK POWER SYSTEMS DYNAMICS AND CONTROL SYMPOSIUM, JULY 25-30, 2022, BANFF, CANADA}
%\rhead{1}

%\fontfamily{phv}\fontseries{b}\fontsize{9}{11}\selectfont

\maketitle
\thispagestyle{fancy}
\pagestyle{fancy}

%\thispagestyle{fancy}
%\pagestyle{fancy}

\begin{abstract}
We develop and test a data-driven and area-based fast frequency control scheme, which rapidly redispatches inverter-based resources to compensate for local power imbalances within the bulk power system. The approach requires no explicit system model information, relying only on historical measurement sequences for the computation of control actions. Our technical approach fuses developments in low-gain estimator design and data-driven control to provide a model-free and practical solution for fast frequency control. Theoretical results and extensive simulation scenarios on a three area system are provided to support the approach.
\end{abstract}

\begin{IEEEkeywords}
renewable energy, frequency control, transmission grid, smart grid
\end{IEEEkeywords}


% \fi
\IEEEpeerreviewmaketitle

\section{Introduction}
\label{Sec:Introduction}

% {\tb
% \textbf{Ongoing:}
% \begin{enumerate}
% \item See how De Persis or Van Waarde papers can be leveraged to design better excitation signal for nonlinear systems
% \item Is it possible for slow load variations to be sufficient on our time-scale of interest?
% \item Do results improve if we use low rank approximation for Hankel matrices before solving the linaer equations?
% \end{enumerate}
% %
% \textbf{Future Ideas:}
% \begin{enumerate}
% \item For application in very large interconnected systems, is it possible to use other measurements instead of frequency?
% \item Incorporate slack variables into the linear equations and solve a regularized QP to determine $\hat{y}(t)$; this follows the ideas of Coulson et al. on improving robustness to measurement noise
% \item MHE disturbance estimator
% \item Improved design of excitation signal to minimize system impact during data collection
% \item Can historical event data be used? Can we piece together trajectories from different times to form one persistently exciting signal.
% \end{enumerate}
% }

Driven by the need to decarbonize the existing power system generation infrastructure \cite{irena2018global}, the transmission grid is increasingly being dominated by inverter-based renewable energy resources (IBRs). Challenges arising from this transition away from traditional generation include larger (and more frequent) frequency deviations, faster frequency dynamics due to reduced system inertia, and increased net load variability \cite{irena2018global,poolla2017optimal}. As a result of this this proliferation of IBRs in the grid, there is an increasing urgency to develop new and faster frequency control methods.

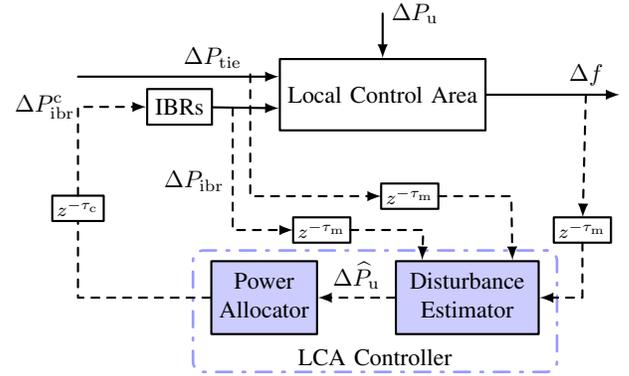
\begin{figure}[ht!]
  \begin{minipage}{0.5\textwidth}
\begin{center}
    \begin{tikzpicture}[auto, scale = 0.6, node distance=2cm,>=latex', every node/.style={scale=0.9},
      blacknode/.style={shape=circle, draw=black, line width=2},
  bluenode/.style={shape=circle, draw=blue, line width=2},
  greennode/.style={shape=circle, draw=green, line width=2},
  rednode/.style={shape=circle, draw=red, line width=2},
  dotted_block/.style={draw=blue!40!white, line width=1pt, dash pattern=on 1pt off 4pt on 6pt off 4pt,
            inner ysep=6mm,inner xsep=4mm, rectangle, rounded corners}
    ]
      \tikzstyle{anch} = [coordinate]
      \tikzstyle{block} = [draw, fill=white, rectangle, 
      minimum height=3em, minimum width=6em]
      \tikzstyle{bigblock} = [draw, fill=white, rectangle, 
      minimum height=5em, minimum width=25em]
      \tikzstyle{wideblock} = [draw, fill=white, rectangle, 
      minimum height=3em, minimum width=9em]
      \tikzstyle{hold} = [draw, fill=white, rectangle, 
      minimum height=1.5em, minimum width=2em]
      \tikzstyle{delayblock} = [draw, fill=white, rectangle, 
      minimum height=1em, minimum width=0.7em, inner sep=2pt, outer sep=0pt]
      \tikzstyle{sum} = [draw, fill=white, circle, minimum size=0.2cm, node distance=1cm, inner sep=0pt, outer sep=0pt]
      \tikzstyle{mynode} = [draw, thick, fill=white, circle]
      \tikzstyle{input} = [coordinate]
      \tikzstyle{cblue}=[circle, draw, thick, fill=white, scale=1]
      \tikzstyle{output} = [coordinate]

      %% Place block
      \node[block] (power_system)  {Local Control Area};
      \node[block, below of=power_system,fill=blue!20, node distance=3cm, xshift=1.25cm, name=] (estimator) {\makecell[c]{Disturbance \\Estimator}};
      
      %\node[hold, left of=estimator,fill=blue!20, node distance=2.5cm] (filter) {\makecell[c]{Detuning \\Filter}};
      
      \node[hold, left of=estimator,fill=blue!20, node distance=3cm] (allocator) {\makecell[c]{Power \\Allocator}};
      \node [dotted_block, fit = (estimator) (allocator),inner ysep=4mm, inner xsep=5mm,xshift=0mm,yshift=-2mm] (controller) {};
      \node at (controller.south) [above, inner sep=1mm,] {LCA Controller};

      \node[hold, left of=power_system, node distance=3cm, yshift=-0.18cm, name=IBRs] {IBRs};

	  %Changed to anchor to remove noise inputss
      \node[anch, above of=estimator, node distance=1.5cm, xshift=0.65cm] (sum2) {};
      \node[anch, above of=estimator, node distance=1cm, xshift=-0.65cm] (sum3) {};
      \node[anch, right of=estimator, node distance=1.7cm] (sum4) {};

      %\node[input, right of=sum2, node distance=0.7cm] (noise_2) {};
      %\node[input, right of=sum3, node distance=0.7cm] (noise_3) {};
      %\node[input, right of=sum4, node distance=0.7cm] (noise_4) {};

	  \node[delayblock, left of=sum2, node distance=1.5cm] (delay2) {\footnotesize {$z^{-\tau_{\mathrm{m}}}$}};
	  \node[delayblock, left of=sum3, node distance=1.5cm] (delay3) {\footnotesize {$z^{-\tau_{\mathrm{m}}}$}};	  
	  \node[delayblock, above of=sum4, node distance=1cm] (delay4) {\footnotesize {$z^{-\tau_{\mathrm{m}}}$}};

      %%Place additional inputs
      \node[input, above of=power_system, node distance=1.2cm, name=load] {};      
      \node[input, left of=power_system, node distance=4.5cm, name=Ptie] {};   
      \node[output, right of=power_system, node distance=3.5cm, name=w] {};   
      
      \node[anch, left of = IBRs, node distance=1.5cm] (tmp1) {};

      %Place IBR command block
      \node[delayblock, below of=tmp1, node distance=1.5cm] (delay_ibr) {\footnotesize {$z^{-\tau_{\mathrm{c}}}$}};

      %% Connect blocks and inputs
      \draw [thick,-latex] (IBRs) -- node[name=p_ibr,pos=0.3] {} ([yshift=-0.3cm]power_system.west);
      %\draw[thick,dashed,-latex] (estimator) -- (filter);
      %\draw[thick,dashed,-latex] (filter) -- (allocator);
      \draw[thick,dashed,-latex] (estimator) -- node[above] {$\Delta \widehat{P}_{\rm u}$} (allocator);

      \draw [thick,-latex] ([yshift=0.4cm]Ptie) -- node[above, pos=0.67, name = p_tie] {$\Delta P_{\mathrm{tie}}$} ([yshift=0.4cm]power_system.west);
      
      \draw [thick, -latex] (load) -- node[pos=0, right] {$\Delta P_{\mathrm{u}}$} (power_system);
      \draw [thick,-latex] (power_system.east) -- node[pos=0.75, name=freq] {$\Delta f$} (w);

      \draw [thick,dashed,-latex] (freq) -- (delay4);
      \draw [thick,dashed,-latex] (delay4) -- (sum4) -- (estimator);

	  \draw [thick,dashed,-latex] (allocator.west) -| node[pos = 0.2, left, xshift=-0.2cm] {} (delay_ibr) |- node[pos=0.64, left, xshift=-0.2cm] {$\Delta P_{\mathrm{ibr}}^{\mathrm{c}}$} (IBRs) ;
	  
	  \draw [thick,dashed, -latex] ([xshift=7cm]p_tie) |- (delay2);
	  \draw [thick,dashed, -latex] (delay2) -- (sum2) -- ([xshift=1cm]estimator.north);
	  
	  \draw [thick,dashed, -latex] (p_ibr) |- node[pos=0.3, left] {$\Delta P_{\mathrm{ibr}}$} (delay3);
	  \draw [thick,dashed, -latex] (delay3) -- (sum3) -- ([xshift=-1cm]estimator.north);

      \end{tikzpicture}
      \end{center}
      \caption{Block diagram of area control structure for each LCA. Dashed lines denote sampled signals.}
      \label{fig:acs02}
\end{minipage}
\end{figure}

 To address the aforementioned challenges, in \cite{ekomwenrenren2021hierarchical} the authors proposed a hierarchical control scheme which coordinates IBRs to provide fast frequency control. In this scheme, the bulk grid is partitioned into geographically small local control areas (LCAs). Within each LCA, a local controller is designed which processes (potentially, delayed) measurements from the LCA to compute updated set-points for local IBRs (Figure \ref{fig:acs02}). The local controller has two key sub-blocks. The \emph{disturbance estimator} detects frequency events (e.g., a large load disturbance and/or generation outage) by computing a real-time estimate $\Delta \widehat{P}_{\rm u}$ of the net unmeasured active power imbalance $\Delta P_{\rm u}$ in the LCA. The \emph{power allocator} then continuously and optimally redispatches the IBRs to correct the imbalance, subject to device limits. In situations where local IBR resources are insufficient, a higher-level coordinating  controller facilitates the optimal transfer of additional power support from neighbouring LCAs. As our focus in this paper will be on the LCA controller design, this coordinating control layer will not be discussed further; see \cite[Sec. III-B]{ekomwenrenren2021hierarchical} for details.

The key to the fast operation of this scheme lies in disturbance estimator, which must quickly produce an accurate estimate of the power imbalance within the LCA based on the available measurements. In \cite{ekomwenrenren2021hierarchical}, the disturbance estimator was designed in a model-based fashion, using a crude second-order lumped dynamic LCA model \cite{anderson1990low}
\begin{equation}\label{Eq:SFR}
\resizebox{0.9025\linewidth}{!}{$
\begin{aligned}
    2H\Delta{\dot{\omega}} &= - \tfrac{1}{R_{\mathrm I}}\Delta \omega + \Delta P_{\mathrm{m}} - \Delta P_{\mathrm{u}} - \Delta P_{\mathrm{tie}} + \Delta P_{\mathrm{ibr,tot}}^{\mathrm{c}}\\
    T_{\rm R}\Delta{\dot{P}_{\mathrm{m}}} &= - \Delta P_{\mathrm{m}} - R_{\rm g}^{-1}(\Delta{\omega} + T_{\mathrm{R}}F_{\mathrm{H}}\Delta{\dot{\omega}})
\end{aligned}
$}
\end{equation}
parameterized by several constants, such as total LCA inertia $H$, total IBR and generator primary control gains $R_{\rm I}$ and $R_{\rm g}$,  aggregated turbine-governor time constant $T_{\rm R}$, and aggregated  high-pressure turbine fraction $F_{\rm H}$. This modelling decision was made for pragmatic reasons, as more accurate high-order system models could be too burdensome to build and maintain in practice \cite{prostejovsky2017tuningless}. Tests in \cite{ekomwenrenren2021hierarchical} demonstrated that this design approach can yield excellent closed-loop control when the model \eqref{Eq:SFR} reasonably captures the LCA dynamics and when the lumped parameters in \eqref{Eq:SFR} are accurately set. However, when either of these conditions fails, control performance (e.g., post-disturbance settling time and overshoot) will indeed deteriorate.

Many modern approaches to frequency control, such as robust optimal control \cite{bevrani2014robust}, model predictive control \cite{venkat2008distributed,ersdal2015model,kohler2017real,monasterios2017low} and the coordinated dispatch of IBRs \cite{ekomwenrenren2021hierarchical}, are of similar vein in that they require accurate model information about the power system, which may difficult to obtain in practice \cite{prostejovsky2017tuningless}. In contrast to model-based approaches, a model-free or data-driven approach would allow for better adaptation to realistic system conditions, resulting in faster disturbance estimation and rejection, and with reduced tuning of parameters required for real-world implementation. Our objective in this paper is to develop such a data-driven frequency controller.

%\john{Things become very messy after this point in the introduction, needs more attention from you.}

%Therefore, it is important to consider data-driven approaches to reduce the dependence of the control scheme on explicit model information. We explore the current literature on such approaches in the subsequent discussion.
%Unfortunately, despite these efforts to minimize the amount and quality of system model information required, in most cases accurate parameter values for use in the models cannot be easily obtained in practice with reasonable effort, or cannot be accurately updated as system conditions change. 

In the literature, various researchers have tackled the problem of power system frequency control using data-driven and learning-based approaches. The most popular approach is based on reinforcement learning, wherein control actions are undertaken to maximize some form of cumulative reward \cite{chen2020model, yu2011stochastic, yan2018data, yan2020multi}. Drawbacks of this approach are that it requires a significant number of data samples, and can be particularly sensitive to the selection of hyper-parameters \cite{recht2019tour}. 
%
%Another approach for designing model-free controllers for frequency control is the use of adaptive-based control techniques such as fuzzy-logic and dynamic programming \cite{dong2016event,mu2017improved,yousef2014load}. However, such techniques are not easily amenable to analytic analysis and there has been reported unexpected instabilities in real-world implementation \cite{anderson2005failures}.
%%
Other researchers take a hybrid approach, fusing elements of data-driven control with sequential system identification (ID) and control \cite{liu2019model,hidalgo2019frequency}. However, since a system identification step is needed to identify an approximate model, the issue of non-trivial parameter tuning still remains.

With the goal of using a smaller number of data samples and circumventing the need of a system identification step, we base our approach here for the design of a frequency controller on recent data-driven predictive control approaches proposed in the literature, which compute control actions without explicitly identifying the system from data \cite{markovsky2008data,coulson2019data,de2019formulas,berberich2020data,van2020data,zhao2021data}. These design techniques are based on a fundamental result \cite{JCW-PR-IM-BLMDM:05} in the behavioral approach to systems and control theory \cite{willems1997introduction}, namely that a single recorded historical trajectory of a controllable LTI system implicitly defines a model of the system, as long as the input sequence is persistently exciting (i.e., sufficiently rich). Authors in \cite{zhao2021data} utilize this technique in a data-driven model predictive control framework to regulate frequency in the power system context. However, in their approach, the net-load demand in the system is assumed available and used as a control input; this information may be difficult to obtain precisely at all times, especially with the added net-load variability and uncertainty in power supply introduced by renewables \cite{Ulbig2014}.

%\john{I am baffled at this point -- why would your discussion of data-driven control approaches not comment on the actual line of data-driven control approaches -- based on behavioral system theory -- on which the entire approach here is based? You need to be citing Coulson, other papers from Dorfler group, De Persis, Markosvsky, other related papers from this area, etc.}

\emph{Contributions:} The main contributions of this paper are summarized as follows. First, in Section \ref{Sec:DataDrivenDisturbanceEstimator} we combine ideas from dynamic estimator theory and behavioural systems theory to design a data-based substitute for the model-based disturbance estimator in \cite{ekomwenrenren2021hierarchical}. Second, in Section \ref{Sec:Application}, we leverage the theory of Section \ref{Sec:DataDrivenDisturbanceEstimator} to propose a novel model-free frequency controller, which achieves fast frequency regulation by rapidly redispatching inverter-based resources to compensate for local power imbalances within the bulk power system. The key advantage of our approach is that it does not rely on a parametric system representation for computing control actions; the LCA system model in Figure \ref{fig:acs02} is replaced by time-series data, which is sufficiently rich in harmonic content to capture the dynamics of the system. This time-series data is then directly used in our hierarchical control framework, without passing through an explicit system identification step. Third and finally, in Section \ref{Sec:CaseStudy}, we extensively validate our approach via simulations on a detailed nonlinear three-area power system presented in Section \ref{Sec:TestSystem}. Across several scenarios \textemdash{} including load increases, heavy-renewable penetration, generation trips, and three-phase faults \textemdash{} we illustrate how our data-driven approach can provide fast and effective frequency control for the bulk grid.

%Furthermore, by incorporating a regularization step via low-rank approximation of the data matrix, we can utilize fewer data samples and potentially lower excitation signal power, as shown in Section \ref{Sec:TestSystem}. In this work, we have used 101 data samples with a measurement sampling time of 0.1s and an excitation signal consisting of a sinusoidal perturbation of 1 MW, with band-limited white noise with power spectral density (PSD) of 0.1.

%The remainder of this paper is organized as follows. Sections \ref{Sec:ControlPreliminaries} and \ref{Sec:DataDrivenDisturbanceEstimator} review the control preliminaries and cover the theory on data-driven disturbance estimator design. The proposed data-driven frequency controller design is highlighted in Section \ref{Sec:DataDrivenDisturbanceEstimator}. In Section \ref{Sec:Application}, we describe how the theory outlined in Section \ref{Sec:ControlPreliminaries} and Section \ref{Sec:DataDrivenDisturbanceEstimator} is adapted for application to the area-based fast frequency control problem of current interest.Section \ref{Sec:TestSystem} and \ref{Sec:CaseStudy} present the nonlinear three-area test system, along with simulation results validating our approach. Section \ref{Sec:Conclusion} concludes the report and outlines future research directions.}

\emph{Notation:} Given vectors or matrices $x_1,\ldots,x_N$ which have equal numbers of columns, $\mathrm{col}(x_1,\ldots,x_N)$ denotes their column concatenation. For a matrix $A$ having full column rank, $A^{\dagger} = (AA^{\sf T})^{-1}A^{\sf T}$ denotes its pseudoinverse.

\section{Control Preliminaries}
\label{Sec:ControlPreliminaries}

Our approach relies on a technical foundation built from two topics in linear control theory: disturbance estimator design, and behavioral systems theory. Section \ref{Sec:DistEst} recalls some preliminaries on disturbance estimator design. Section \ref{Sec:LowGain} provides additional insights and assumptions that simplify the disturbance estimator problem introduced in Section \ref{Sec:DistEst}. Section \ref{Sec:Behavioral} introduces the basic facts from behavioral systems theory that enable data-driven prediction based on historical system trajectories. 

\subsection{Background on Disturbance Estimators}
\label{Sec:DistEst}

Consider the discrete-time linear time-invariant (LTI) model
\begin{equation}\label{Eq:LTI}
\begin{aligned}
x(t+1) &= Ax(t) + Bu(t) + B_d d(t)\\
y(t) &= Cx(t) + Du(t)
\end{aligned}
\end{equation}
with time index $t \in \integer_{\geq 0}$, state $x(t) \in \real^n$, control input $u(t) \in \real^m$, disturbance input $d(t) \in \real^{q}$, and output $y(t) \in \real^{p}$. We assume that $d(t)$ is a \emph{constant} but unknown disturbance signal and that $q \leq p$. The problem of \emph{asymptotic disturbance estimation} is to design a causal system which processes $(u(t),y(t))$ to produce an estimate $\hat{d}(t)$ of $d(t)$ at time $t$, which satisfies $\lim_{t\to\infty}(\hat{d}(t) - d(t)) = 0$ irrespective of the initial conditions.

This problem admits a standard model-based solution via the design of a Luenberger observer for the extended model
\begin{equation}\label{Eq:LTIExtended}
\begin{aligned}
\xi(t+1) &= \underbrace{\begin{bmatrix}
A & B_d\\
0 & I_q
\end{bmatrix}}_{\define \mathcal{A}}\xi(t) + \underbrace{\begin{bmatrix}B\\ 0\end{bmatrix}}_{\define \mathcal{B}}u(t)\\
y(t) &= \underbrace{\begin{bmatrix}C & 0\end{bmatrix}}_{\define \mathcal{C}}\xi(t)  + \underbrace{D}_{\define \mathcal{D}}u(t).
\end{aligned}
\end{equation}
where $\xi(t) = \mathrm{col}(x(t),d(t))$ is the extended state vector. In \eqref{Eq:LTIExtended}, the dynamic disturbance model $d(t+1) = d(t)$ with unknown initial condition is equivalent to the constant disturbance in the model \eqref{Eq:LTI}. A standard predictive-type Luenberger observer for \eqref{Eq:LTIExtended} can now be designed as
\begin{equation}\label{Eq:Luenberger1}
\begin{aligned}
\hat{\xi}(t+1) &= \mathcal{A}\hat{\xi}(t) + \mathcal{B}u(t) - \mathcal{L}(\hat{y}(t)-y(t))\\
\hat{y}(t) &= \mathcal{C}\hat{\xi}(t)  + Du(t)
\end{aligned}
\end{equation}
where $\mathcal{L} \in \real^{(n+q)\times p}$ is the estimator gain to be designed. With $\tilde{\xi}(t) = \xi(t) - \hat{\xi}(t)$ denoting the estimation error, the error dynamics are computed to be
\begin{equation}\label{Eq:ErrorDynamics}
\tilde{\xi}(t+1) = (\mathcal{A}-\mathcal{L}C)\tilde{\xi}(t).
\end{equation}
It is a standard result that there exists $\mathcal{L}$ such that the error dynamics \eqref{Eq:ErrorDynamics} are exponentially stable if and only if $(A,C)$ is detectable and if the plant \eqref{Eq:LTI} has no transmission zeros at $z = 1$ on the $d \mapsto y$ channel (see, e.g., \cite[Chp. 23]{levine2018control}. Under these conditions, the estimator \eqref{Eq:Luenberger1} solves the asymptotic disturbance estimation problem, with the estimate $\hat{d}(t)$ being obtained from the second block component of $\hat{\xi}(t)$.

\subsection{Low-Gain Disturbance Estimation}
\label{Sec:LowGain}

The design of the estimator gain $\mathcal{L}$ above requires the solution of a standard multivariable stabilization problem for the detectable pair $(\mathcal{A},\mathcal{C})$ of extended plant matrices. In particular then, the design requires knowledge of the entire system matrix $A$, describing the internal state dynamics of the model. Under additional assumptions, this design requirement can be significantly relaxed and made more practical.

First, we will assume that $A$ is Schur stable, i.e., that the plant \eqref{Eq:LTI} is internally exponentially stable. This assumption would be satisfied by, for example, a linearized power system model describing frequency dynamics and incorporating typical primary controllers. Under this assumption $(A,C)$ is automatically detectable, and the transmission zero condition reduces to the condition that $G_{d}(1) \in \real^{p \times q}$ have full column rank, where $G_{d}(z) = C(zI-A)^{-1}B_d$ denotes the transfer matrix on the $d \to y$ channel. In this case, if one seeks an estimator gain of the form 
\[
\mathcal{L} = \begin{bmatrix}0 \\ \varepsilon L\end{bmatrix}
\]
for some $L \in \real^{q \times p}$ and $\varepsilon > 0$, then \eqref{Eq:Luenberger1} reduces to
\begin{subequations}\label{Eq:ReducedEstimator}
\begin{align}
\label{Eq:ReducedEstimator1}
\hat{x}(t+1) &= A\hat{x}(t) + Bu(t) + B_d\hat{d}(t)\\
\label{Eq:ReducedEstimator2}
\hat{y}(t) &= C\hat{x}(t) + Du(t)\\
\label{Eq:ReducedEstimator3}
\hat{d}(t+1) &= \hat{d}(t) - \varepsilon L(\hat{y}(t)-y(t)).
\end{align}
\end{subequations}
The estimator \eqref{Eq:ReducedEstimator} admits the following simple interpretation. The first two equations \eqref{Eq:ReducedEstimator1}--\eqref{Eq:ReducedEstimator2} perform an open-loop simulation of the plant, using the control input $u(t)$ and the \emph{estimated} disturbance $\hat{d}(t)$ to produce an estimate $\hat{y}(t)$ for the output. Given $\hat{y}(t)$ and the true output measurement $y(t)$, the equation \eqref{Eq:ReducedEstimator3} updates the disturbance estimate. Crucial to our data-driven development to follow is the following observation: the only purpose of \eqref{Eq:ReducedEstimator1}--\eqref{Eq:ReducedEstimator2} is to produce the output estimate $\hat{y}(t)$; the state estimate $\hat{x}(t)$ is irrelevant for our purposes. 

The following result, which will not be proven here, guarantees that the design of such an estimator is possible.

\smallskip

\begin{proposition}[\bf Low-Gain Disturbance Estimator Design]\label{Prop:LowGain}
Consider the disturbance estimator \eqref{Eq:ReducedEstimator} for the plant \eqref{Eq:LTI}. Assume that $A$ is Schur stable, that $G_{d}(1) = C(I_n-A)^{-1}B_d$ has full column rank, and set $L = G_{d}(1)^{\dagger}$. Then there exists $\varepsilon^{\star} > 0$ such that for all $\varepsilon \in (0,\varepsilon^{\star})$, the estimator \eqref{Eq:ReducedEstimator} solves the asymptotic disturbance estimation problem. 
\end{proposition}

% \begin{pfof}{Theorem \ref{Prop:LowGain}}
% A formal proof is not difficult; here we only sketch the main idea. The error dynamics \eqref{Eq:ErrorDynamics} now become
% \[
% \begin{aligned}
% \tilde{\xi}_1(t+1) &= A\tilde{\xi}_1(t) + B_{d}\tilde{\xi}_2(t)\\
% \tilde{\xi}_{2}(t+1) &= - \varepsilon LC \tilde{\xi}_1(t) + \tilde{\xi}_{2}(t).
% \end{aligned}
% \]
% For small values of $\varepsilon$, $\tilde{\xi}_2$ barely changes from step to step. Since $A$ is Schur stable, it follows that $\tilde{\xi}_1$ will converge to the quasi steady-state value $\tilde{\xi}_1 = (I_n-A)^{-1}B_d\tilde{\xi}_2$, and hence the $\tilde{\xi}_2$ equation reduces to
% \[
% \begin{aligned}
% \tilde{\xi}_2(t+1) &= - \varepsilon LC (I_n-A)^{-1}B_d\tilde{\xi}_2(t) + \tilde{\xi}_{2}(t)\\
% &= (I_q - \varepsilon LG(1))\tilde{\xi}_2(t)\\
% &= (1 - \varepsilon)\tilde{\xi}_2(t)
% \end{aligned}
% \]
% since $L = G(1)^{\dagger}$. The latter system is always stable for sufficiently small $\varepsilon > 0$, which completes the proof.
% \end{pfof}

\smallskip

Proposition \ref{Prop:LowGain} states that with the simple estimator gain selection $L = G_{d}(1)^{\dagger}$, one can always obtain a stable estimator by starting $\varepsilon > 0$ at a small value and then tuning. The quantity $G_{d}(1)$ is known as the DC gain, and can be obtained directly from recorded data (See Section \ref{Sec:DataDrivenDisturbanceEstimator} for details). The above result also applies without changes to modified estimator
\begin{subequations}\label{Eq:CurrentEstimator}
\begin{align}
\label{Eq:CurrentEstimator1}
\hat{x}(t+1) &= A\hat{x}(t) + Bu(t) + B_d\hat{d}(t)\\
\label{Eq:CurrentEstimator2}
\hat{y}(t) &= C\hat{x}(t) + Du(t)\\
\label{Eq:CurrentEstimator3}
\hat{d}(t) &= \hat{d}(t-1) - \varepsilon L(\hat{y}(t)-y(t)).
\end{align}
\end{subequations}
which uses the most recent measurement $y(t)$ to compute $\hat{d}(t)$ as opposed to \eqref{Eq:ReducedEstimator3} which uses $y(t-1)$ to compute $\hat{d}(t)$.

\subsection{Primer on Behavioral Systems Theory}
\label{Sec:Behavioral}

The model defined by \eqref{Eq:LTI} is a parametric representation of a LTI system. In the setting of behavioral systems theory, a LTI system is instead interpreted as defining an implicit constraint on the vector spaces of input and output signals. Our treatment here will be very minimal; see \cite{willems1997introduction} for a detailed introduction.

Let $\sigma$ denote the signal shift operator defined by $(\sigma x)(t) = x(t+1)$. The \emph{behavior} $\mathscr{B}$ of \eqref{Eq:LTI} is defined as the set of all possible input-output sequences which are consistent with the model
\[
\begin{aligned}
\mathscr{B} = &\left\{(u,d,y) \in (\real^{m+q+p})^{\integer_{\geq 0}}\,\,:\,\,\exists x\in(\real^{n})^{\integer_{\geq 0}}\,\,\text{s.t.}\right.\\
& \qquad\left.\sigma x = Ax + Bu + B_dd,\,\, y=Cx+Du\right\}.
\end{aligned}
\]
The smallest possible state dimension consistent with $\mathscr{B}$ is called the \emph{order} of the system, and is denoted by $n(\mathscr{B})$. The \emph{lag} of $\mathscr{B}$, denoted by $\ell(\mathscr{B})$ is the smallest integer $\ell$ such that the matrix $\mathcal{O}_{\ell} = \mathrm{col}(C,CA,\ldots,CA^{\ell-1})$ has rank $n(\mathscr{B})$. As additional notation, we let $\mathscr{B}_{N}$ denote the restriction of the behaviour to trajectories of length $N \in \integer_{\geq 1}$.

The following notion is essential. Let $T \in \integer_{\geq 1}$ and let $z \in (\real^m)^{T}$ be the length $T$ signal
\[
z = \mathrm{col}(z(1),\ldots,z(T)).
\]
We say $z$ is \emph{persistently exciting of order $L$} if the \emph{Hankel matrix}
\[
\mathscr{H}_{L}(z) = \begin{bmatrix}z(1) & \cdots & z(T-L+1)\\
\vdots & \ddots & \vdots\\
z(L) & \cdots & z(T)
\end{bmatrix} \in \real^{mL \times (T-L+1)}
\]
has full row rank. The idea is that if $z$ is persistently exciting, then it is both sufficiently long and sufficiently rich in variation.

Suppose that, from the plant \eqref{Eq:LTI}, we have collected $T \in \integer_{\geq 1}$ samples of input and output data
\begin{equation}\label{Eq:HistoricalData}
\begin{aligned}
u_{\rm d} &= \mathrm{col}(u(1),\ldots,u(T)) \in (\real^{m})^{T}\\
d_{\rm d} &= \mathrm{col}(d(1),\ldots,d(T)) \in (\real^{q})^{T}\\
y_{\rm d} &= \mathrm{col}(y(1),\ldots,y(T)) \in (\real^{p})^{T}.
\end{aligned}
\end{equation}
A key result known as the \emph{fundamental lemma} \cite{de2019formulas} provides a characterization of \emph{all possible} length $N$ system trajectories in terms of recorded data.

\begin{lemma}[\bf Fundamental Lemma]\label{Lem:Fundamental}
Assume that $\mathscr{B}$ is controllable and that $\mathrm{col}(u_{\rm d},d_{\rm d})$ is persistently exciting of order $N + n(\mathscr{B})$. Then any possible length $N$ trajectory $(u,d,y)$ of $\mathscr{B}$ can be represented as
\[
\begin{bmatrix}
\mathscr{H}_{N}(u_{\rm d})\\
\mathscr{H}_{N}(d_{\rm d})\\
\mathscr{H}_{N}(y_{\rm d})
\end{bmatrix}g = \begin{bmatrix}u \\ d \\ y\end{bmatrix}
\]
for some vector $g \in \real^{T-N+1}$.
\end{lemma}

\medskip

Put simply, \emph{all possible} system trajectories of length $N$ can be expressed as a linear combination of the columns of the given matrix, which consists purely of recorded historical data. 

Lemma \ref{Lem:Fundamental} enables data-driven simulation of a LTI system \cite{de2019formulas}. To obtain a unique response, a form of initilization is required. This is done through an initial prefix trajectory of length $T_{\rm ini}$ is required for initialization; the technical requirement is that $T_{\rm ini} \geq \ell(\mathscr{B})$. Let $(u_{\rm p},d_{\rm p},y_{\rm p})$ be this prefix trajectory. Let
\[
\begin{bmatrix}
U_{\rm p}\\
U_{\rm f}
\end{bmatrix} = \mathscr{H}_{T_{\rm ini}+N}(u_{\rm d}), \qquad 
\]
be a corresponding partitioning of the Hankel matrices, with similar partitionings for the data $d_{\rm d}$ and $u_{\rm d}$. Leveraging Lemma \ref{Lem:Fundamental}, and assuming that $u_{\rm d}$ is persistently exciting of order $T_{\rm ini} + N + n(\mathscr{B})$, we have the following: given length $N$ inputs $u = (u(T_{\rm ini}+1),\ldots,u(T_{\rm ini}+N)$ and $d = (d(T_{\rm ini}+1),\ldots,d(T_{\rm ini}+N)$, one may uniquely compute the response $y = (y(T_{\rm ini}+1),\ldots,y(T_{\rm ini} + N))$ by solving the system of linear equations
\begin{equation}\label{Eq:DataDrivenSimulation}
\begin{bmatrix}
U_{\rm p}\\
D_{\rm p}\\
Y_{\rm p}\\
U_{N}\\
D_{N}\\
Y_{N}
\end{bmatrix}g = \begin{bmatrix}
u_{\rm p}\\
d_{\rm p}\\
y_{\rm p}\\
u\\
d\\
y
\end{bmatrix}
\end{equation}
in the unknowns $g$ and $y$. Specifically, one solves the first five block equations for $g$, then substitutes into the last equation to compute the response $y$.

\section{Data-Driven Disturbance Estimators}
\label{Sec:DataDrivenDisturbanceEstimator}

We now combine the disturbance estimation ideas Section \ref{Sec:LowGain} with the behavioral systems theory from Section \ref{Sec:Behavioral} to develop a data-driven version of the disturbance estimator \eqref{Eq:CurrentEstimator}.\footnote{A data-driven version of the estimator \eqref{Eq:ReducedEstimator} can be similarly obtained by modifying the results that follow.} As in Section \ref{Sec:LowGain}, we assume that $A$ is Schur stable and that $G_{d}(1)$ has full column rank.

Recall that the idea behind the estimator  \eqref{Eq:CurrentEstimator} is to obtain a prediction $\hat{y}(t)$ for $y(t)$ using \eqref{Eq:CurrentEstimator1}--\eqref{Eq:CurrentEstimator2}, and then to use $\hat{y}(t)$ and the current measurement $y(t)$ to update $\hat{d}(t)$ using \eqref{Eq:CurrentEstimator3}. Translating this to a data-driven framework, our approach will be to use \eqref{Eq:DataDrivenSimulation} in place of \eqref{Eq:CurrentEstimator1}--\eqref{Eq:CurrentEstimator2} to generate the estimate $\hat{y}(t)$, and then to verbatim use \eqref{Eq:CurrentEstimator3}. Suppose that we have collected historical data as in \eqref{Eq:HistoricalData}, and that the inputs $\mathrm{col}(u_{\rm d},d_{\rm d})$ are persistently exciting of order $T_{\rm ini} + 1 + n(\mathscr{B})$ with $T_{\rm ini} \geq \ell(\mathscr{B})$. At time step $t$, we construct the vectors of recent past data
\[
\begin{aligned}
\hat{y}_{\rm p} &= \mathrm{col}(\hat{y}(t-T_{\rm ini}),\ldots,\hat{y}(t-1))\\
u_{\rm p} &= \mathrm{col}(u(t-T_{\rm ini}),\ldots,u(t-1))\\
\hat{d}_{\rm p} &= \mathrm{col}(\hat{d}(t-T_{\rm ini}),\ldots,\hat{d}(t-1)).
\end{aligned}
\]
Following \eqref{Eq:DataDrivenSimulation}, we formulate and solve the system of equations
\begin{equation}\label{Eq:FirstEstimator}
\begin{bmatrix}
U_{\rm p} \\ D_{\rm p} \\ Y_{\rm p} \\ U_{\rm f} \\  D_{\rm f} \\ Y_{\rm f}
\end{bmatrix}g = \begin{bmatrix}
u_{\rm p}\\
\hat{d}_{\rm p}\\
\hat{y}_{\rm p}\\
u(t)\\
\hat{d}(t)\\
\hat{y}(t)
\end{bmatrix} \quad \Rightarrow \quad \hat{y}(t) = \underbrace{Y_{\rm f}\begin{bmatrix}
U_{\rm p} \\ D_{\rm p} \\ Y_{\rm p} \\ U_{\rm f} \\  D_{\rm f}
\end{bmatrix}^{\dagger}}_{\define \mathcal{M}}\begin{bmatrix}
u_{\rm p}\\
\hat{d}_{\rm p}\\
\hat{y}_{\rm p}\\
u(t)\\
\hat{d}(t)\\
\end{bmatrix}.
\end{equation}
Note that \eqref{Eq:FirstEstimator} allows us to compute an estimate $\hat{y}(t)$ for $y(t)$ using only recorded historical data, recently computed online variables $u_{\rm p}$, $\hat{d}_{\rm p}$, $\hat{y}_{\rm p}$, and the current values $u(t)$, and $\hat{d}(t)$ of the input and disturbance estimates. In fact, we do not even require $\hat{d}(t)$ to evaluate the above, since due to strict causality of \eqref{Eq:LTI} on the $d \mapsto y$ channel, the final column of the matrix $\mathcal{M}$ is identically zero. Our data-driven disturbance estimator is therefore
\begin{subequations}\label{Eq:DataDrivenCurrentEstimator}
\begin{align}
\label{Eq:DataDrivenCurrentEstimator1}
\hat{y}(t) &= \mathcal{M}\mathrm{col}(
u_{\rm p},
\hat{d}_{\rm p},
\hat{y}_{\rm p},
u(t),
\vzeros[q])\\
\label{Eq:DataDrivenCurrentEstimator2}
\hat{d}(t) &= \hat{d}(t-1) - \varepsilon L(\hat{y}(t) - y(t)).
\end{align}
\end{subequations}

% \smallskip

% \begin{lemma}[\bf Strong Causality and Zero Blocks]\label{Lem:Causal}
% The final column of the matrix $\mathcal{M}$ is identically zero.
% \end{lemma}

% \begin{pfof}{Lemma \ref{Lem:Causal}}
% {\tb \bf Etinosa will prove this result.}
% \end{pfof}

%\smallskip

In the ideal case where measurements are collected without noise, the results produced by the data-driven estimator \eqref{Eq:DataDrivenCurrentEstimator} are identical to the results produced by the model-based estimator \eqref{Eq:CurrentEstimator}, since \eqref{Eq:DataDrivenCurrentEstimator1} is a data-based representation of \eqref{Eq:CurrentEstimator1}--\eqref{Eq:CurrentEstimator2}. Therefore \eqref{Eq:DataDrivenCurrentEstimator} is a data-based substitute for \eqref{Eq:CurrentEstimator}.

%The next result confirms that 

% \smallskip

% \begin{theorem}[\bf Data-Driven Disturbance Estimator]\label{Thm:DataDisturb}
% {\tb \bf Etinosa will write this theorem.}
% \end{theorem}

% \begin{pfof}{Theorem \ref{Thm:DataDisturb}}
% {\tb \bf Etinosa will prove this theorem.}
% \end{pfof}

%\smallskip

The final issue to address concerns the tuning of the estimator gain $L$ in \eqref{Eq:DataDrivenCurrentEstimator}. Proposition \ref{Prop:LowGain} provides the tuning suggestion $L = G_{d}(1)^{\dagger}$. While $G_{d}(1)$ could be obtained empirically from repeated step response experiments, it can also be obtained directly from the exact same historical data used to construct the matrix $\mathcal{M}$ in \eqref{Eq:DataDrivenCurrentEstimator}. The following result is an adaptation of \cite[Thm. 4.1]{GB-MV-JC-ED:21}.

\smallskip

\begin{lemma}[\bf DC Gain From Trajectory Data]\label{Lem:DCGainFromData}
Consider the previously defined historical data $(u_{\rm d},y_{\rm d},d_{\rm d})$ and define
\[
\begin{aligned}
y_{\rm d}^{\rm diff} &= (y_{\rm d}(2)-y_{\rm d}(1),\ldots,y_{\rm d}(T)-y_{\rm d}(T-1)) \in (\real^{p})^{T-1}\\
u_{\rm d}^{\rm diff} &= (u_{\rm d}(2)-u_{\rm d}(1),\ldots,u_{\rm d}(T)-u_{\rm d}(T-1)) \in (\real^{m})^{T-1}\\
\end{aligned}
\]
with associated Hankel matrices $Y^{\rm diff} = \mathscr{H}_{\ell(\mathscr{B})}(y_{\rm d}^{\rm diff})$ and $U^{\rm diff} = \mathscr{H}_{\ell(\mathscr{B})}(u_{\rm d}^{\rm diff})$. Then
\[
G_{d}(1) = Y_{\rm f}\begin{bmatrix}Y^{\rm diff}\\
U^{\rm diff}\\
U_{\rm p}\\
D_{\rm p}
\end{bmatrix}^{\dagger}\begin{bmatrix}
0\\
0\\
0\\
I_{q}
\end{bmatrix}.
\]
\end{lemma}

\smallskip

Combining Lemma \ref{Lem:DCGainFromData} with the tuning $L = G_{d}(1)^{\dagger}$, the disturbance estimator \eqref{Eq:DataDrivenCurrentEstimator} provides a \emph{completely model-free} solution to the asymptotic disturbance estimation problem; the only required tuning is the single scalar parameter $\varepsilon \in (0,1)$. 

\section{Data-Driven Fast Frequency Control using Inverter-Based Resources}
\label{Sec:FastFreqControl}

We now detail the application of our data-driven disturbance estimation methods to fast frequency control using IBRs. First, in Section \ref{Sec:Application} we describe how the theory outlined in Section \ref{Sec:ControlPreliminaries} and Section \ref{Sec:DataDrivenDisturbanceEstimator} is adapted for application to the area-based fast frequency control problem described in Section \ref{Sec:Introduction}. Section \ref{Sec:TestSystem} and \ref{Sec:CaseStudy} present the nonlinear three-area test system, along with simulation results validating our approach on several scenarios including load increases, generation trips, and three-phase faults.

\subsection{Application of the Theory for Multi-Area Frequency Control}
\label{Sec:Application}

Consider a large interconnected power system, partitioned into many small LCAs, each of which contains IBRs which are dispatchable within specified limits. The model \eqref{Eq:LTI} will be taken as describing the linearized frequency dynamics of an LCA around a dispatch point of interest. This model is \emph{unknown}, and may have an arbitrarily high number of states, describing, e.g., the electromechanical dynamics and control systems of synchronous generators, wind turbines, load dynamics, and so forth. 

Due to the small spatial scale of each LCA, the measured frequency is roughly uniform within the LCA, even during transient conditions (modulo, e.g., intra-area modes). Analogously, the effect of a power imbalance within the LCA on the frequency is approximately independent of the specific location of the imbalance within the LCA. Thus, to a good approximation, power imbalance influences the frequency of each LCA in a lumped fashion. As a result of these observations, for the model \eqref{Eq:LTI} we make the following selections for inputs and outputs. The measurement $y(t) = \Delta f(t) \in \real$ will be a single local measurement of frequency deviation\footnote{An average or weighted average of frequency measurements from across the LCA may also be used.}; this further implies that $D = 0$. The disturbance $d(t) = \Delta P_{\rm u} \in \real$ will model aggregate \emph{unmeasured} generation-load imbalance within the LCA; this is the disturbance we wish to quickly compensate through real-time redispatch of IBRs. The control input $u(t) = \Delta P_{\mathrm{ibr,tot}}(t)$ will model the sum of all IBR power set-points, again relative to scheduled dispatch values. As disturbances and generation are effectively lumped, we further assume that $B = B_{d}$, i.e., disturbance and control signals enter through the same channel. All power flows to neighboring LCAs are assumed to be measured, and thus the total tie flow $\Delta P_{\mathrm{tie}}(t)$ out of the area is considered as a \emph{measurable} disturbance, and is lumped with the control signal $u(t)$.

 The procedure used for collecting historical data will be described in the next section. Due to the assumptions that $B = B_{d}$ and $D = 0$, in developing the data-driven estimator of Section \ref{Sec:DataDrivenDisturbanceEstimator}, it suffices to collect historical data $(u_{\rm d},y_{\rm d})$ in the absence of significant unmeasured disturbances, i.e., during normal system conditions. In this case, the estimator \eqref{Eq:DataDrivenCurrentEstimator1} becomes
\begin{subequations}\label{Eq:IBREstimator}
\begin{align}
\Delta \hat{f}(t) &= Y_{\rm f}\begin{bmatrix}
U_{\rm p} \\ Y_{\rm p} \\ U_{\rm f}
\end{bmatrix}^{\dagger}\begin{bmatrix}
\Delta P_{\rm ibr,tot,p} - \Delta P_{\rm tie,p} + \Delta\hat{P}_{\rm u,p}\\
\Delta\hat{f}_{\rm p}\\
0\\
\end{bmatrix}\\
\Delta\hat{P}_{\rm u}(t) &= \Delta\hat{P}_{\rm u}(t-1) - \varepsilon \frac{1}{G_{d}(1)}(\Delta\hat{f}(t) - \Delta f(t)).
\end{align}
\end{subequations}
and the gain computation of Lemma \ref{Lem:DCGainFromData} reduces to
\[
G_{d}(1) = Y_{\rm f}\begin{bmatrix}Y^{\rm diff}\\
U_{\rm p}\\
\end{bmatrix}^{\dagger}\begin{bmatrix}
0\\
1
\end{bmatrix} \in \real.
\]
The algorithm \eqref{Eq:IBREstimator} estimates the load unmeasured imbalance. To compensate the imbalance, the total change in IBR power set-points required within the LCA is updated as
\[
\Delta P_{\rm ibr,tot}(t+1) = \Delta \hat{P}_{\rm u}(t). 
\]
Finally, to compute the set-points for individual IBRs within each LCA, the total set-point change $\Delta P_{\rm ibr,tot}(t)$ is allocated amongst the IBRs. This can be done via the optimal constrained allocation method presented in \cite{ekomwenrenren2021hierarchical}, or alternatively, the total can be allocated to individual IBRs based on participation factors and then saturated to respect the device power limits; the former approach is used here.

\subsection{Description of Test System and Data Collection}
\label{Sec:TestSystem}

The three-LCA test system under consideration is shown in Figure \ref{fig:cpps01}, where each of the individual areas is modified based on the IEEE 3-machine 9-bus system \cite{delavari2018simscape}. In the modified model, two dispatchable IBRs are added to each LCA to facilitate  fast frequency control. For primary frequency support, each IBR unit has a 5\% droop curve on its respective base power. Doubly-fed induction generator (DFIG) wind turbine systems replace two traditional synchronous generators (SGs) in Area 1, and replace one SG each in Area 2 and Area 3. One static var compensator (SVC) has been added to both Area 1 and Area 2, while two SGs in Area 3 are replaced with two IBRs having the same ratings. The pre-disturbance generation data and unit capacity limits for the system can be found in Table \ref{tab:GenInfo}.

\begin{table}[!ht]
    \caption{Generator and IBR Data.}
    %\caption{SFR model parameters for LCA estimator design.}
    \begin{center}
    \scalebox{1}{
    \begin{tabular}{|c|c|c|c|}
        \hline
        Node & Gen. ID & Rating (MVA) & Dispatch (MW)\\
        \hline
        1 & WT1 & 142.2 & 72.24 \\
        \hline
        1 &	IBR2 & 50 & 15\\
        \hline
        3 &	G1  &	192 & 126\\
        \hline
        3 &	IBR1 & 50 & 25\\
        \hline
        5, 11, 17 &	WT2, WT3, WT4 &	142.2 & 85\\
        \hline
        7 &	G2 & 247.5 & 71.99\\
        \hline
        7 &	IBR4 & 50 & 20\\
        \hline
        9 &	G3  &	192 & 133\\
        \hline
        11 & IBR3 & 50 & 10\\
        \hline
         13 & G4 & 247.50 & 72.24 \\
        \hline
        13 & IBR6 & 50 & 5\\
        \hline
        15 & G5 & 192 & 128\\
        \hline
        17 & IBR5 & 50 & 30\\
    \hline
    \end{tabular}
    }
    \end{center}
    \label{tab:GenInfo}
\end{table}

\begin{figure}[ht!]
\centering
\includegraphics[width=1\columnwidth]{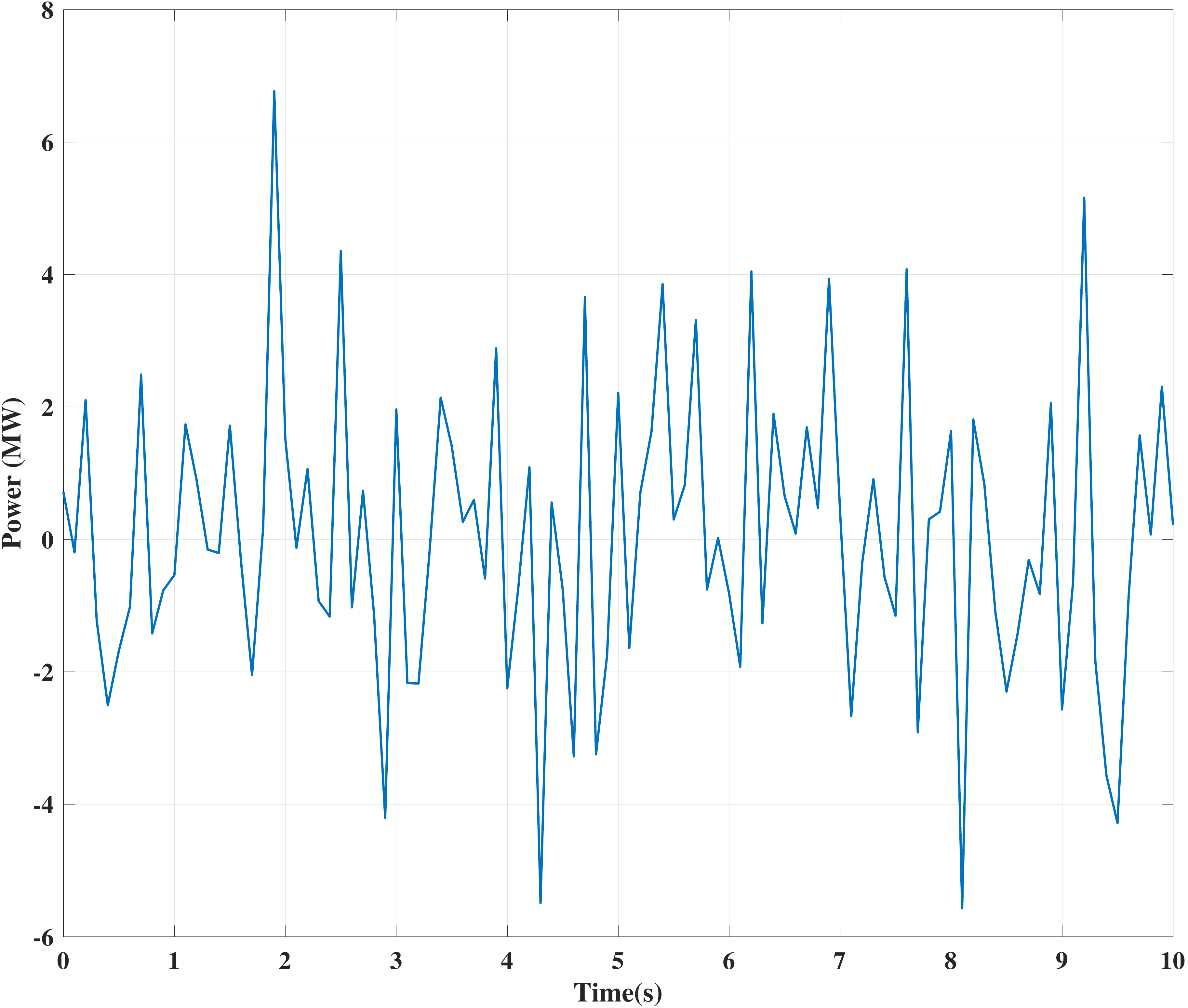}
\caption{Persistently exciting IBR set-point change for data collection phase.}
\label{fig:ExcitationSignal}
\end{figure}

The next step in implementing the data-driven disturbance estimator \eqref{Eq:IBREstimator} is the one-time collection of measurements from the system in response to a sufficiently rich input. To this end, we assume that the IBRs will be used to excite the system for the data collection phase. As described in Section \ref{Sec:DataDrivenDisturbanceEstimator}, the input signal should be persistently exciting of sufficient high order. For the purposes of these tests, the set-point change
\begin{equation}\label{Eq:IBRPerturb}
\Delta P_{\rm ibr}(t) = \sin(12\pi t) + w(t)
\end{equation}
in units of MW was provided to each IBR in each LCA. The signal, plotted in Figure \ref{fig:ExcitationSignal}, consists of a sinusoidal perturbation of 1 MW, with band-limited white noise $w(t)$ with power spectral density (PSD) of 0.1. Note that the amplitude of the perturbing IBR signal is relatively small compared to the overall generation/demand in the system ($\approx 800$ MW).\footnote{Further investigation into the design of practical input signals for data collection is deferred to future work, but see \cite{CDP-PT:21,HJVW:22} for recent theoretical results.}

For each LCA, measurements of frequency deviation and net inter-area power flow were collected for 10 seconds with a sampling period of 0.1s, leading to a total of $T = 101$ total historical data points for each LCA. The length of recent past data used in \eqref{Eq:IBREstimator} was $T_{\rm ini} = 7$, which is a minimal value obtained by starting from a large value and reducing until performance degradation was first observed in simulation. The controller gain $\varepsilon$ in \eqref{Eq:IBREstimator} was set via tuning at $\varepsilon = 0.1$.

Finally, to improve the numerical conditioning of the matrix inversion step in \eqref{Eq:IBREstimator}, the matrix $\mathrm{col}(U_{\rm p},Y_{\rm p},U_{\rm f})$ is replaced by a low-rank approximation thereof, obtained by computing the singular value decomposition and retaining only the first three dominant singular values and vectors \cite{golub2013matrix}. The interpretation of this regularization step is that retaining only a small number of singular values takes into account only the
most dominant sub-behaviours, and therefore removes the effects of unimportant fast dynamics embedded in the data \cite{coulson2019data}.

%That is, we first obtain a low-rank matrix approximation to $col(U_p,D_p,Y_p,U_f,D_f,Y_f)$, before solving the linear equations in Eq. \ref{Eq:FirstEstimator}, via singular value decomposition (SVD), which gives the best low-rank approximation of the data matrix .  

\subsection{Case Studies}
\label{Sec:CaseStudy}

%\john{Did you not want a simple test illustrating the effect of tuning $\varepsilon$?}

We consider {five} disturbance scenarios to illustrate the effectiveness of our data-driven fast frequency controller design. The scenarios are:
\begin{enumerate}
\item[(i)] two large step load changes of different sizes in one area, the first being small enough to compensate with local IBR resources, while the second requires active power support from IBRs in adjacent areas;
\item[(ii)] two simultaneous disturbances, consisting of a step load change and a rate-limited change in the wind power system; the test system in this case is modified to have reduced inertia and increased wind turbine penetration;
\item[(iii)] the loss of a generator;
\item[(iv)] a symmetric three phase-to-ground fault;
\item[(v)] {a step load change in a larger five-area test system.} 
\end{enumerate}

%\john{I think we need to decide where, and to what extent, we will describe the framework and assumptions of the other paper.}

In all scenarios, we consider the hierarchical fast frequency control architecture proposed in \cite{ekomwenrenren2021hierarchical}, and compare the model-based disturbance estimator of that work against the data-driven disturbance estimator presented in this paper.  a comparison of the model-based scheme and an AGC-type scheme can be found in \cite{ekomwenrenren2021hierarchical}, where it was shown that our model-based redispatch can significantly outperform a more traditional AGC-type implementation, and thus we compare only against the model-based scheme. The performance is also compared against a baseline case without any supplementary control scheme, wherein frequency support is provided only through primary droop control action of both generators and IBRs. All simulations are performed with measurement and control signal delays of 300ms, representing worst-case wide-area communication delays, and with white noise of standard deviation $10^{-6}$ (resp. $2\times 10^{-2}$) added to the frequency (resp. inter-area power flow) p.u. measurements.

\begin{figure}
%\begin{minipage}{0.55\textwidth}
\centering
\includegraphics[width=1\columnwidth]{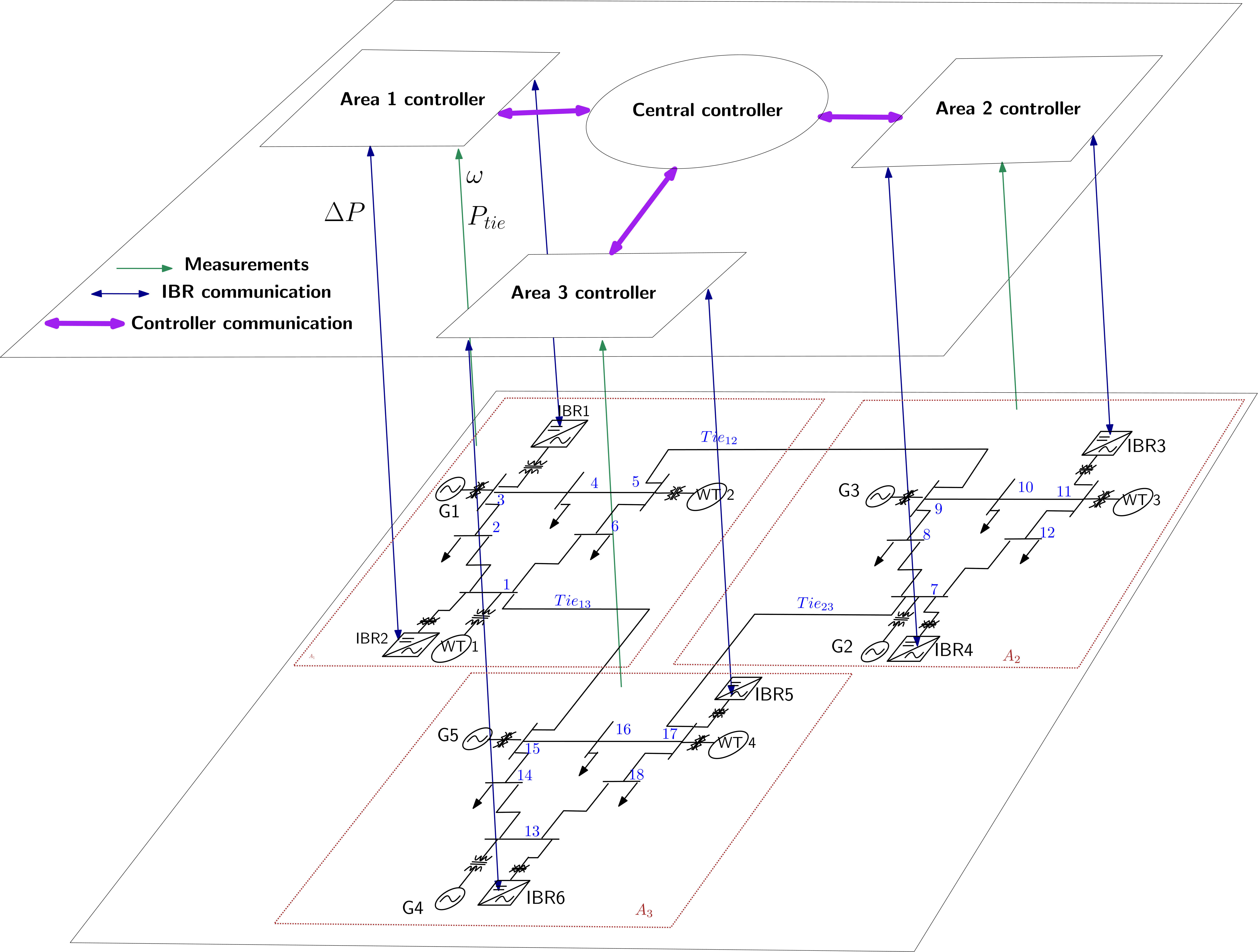}
\caption{Cyber-physical system illustrating frequency control approach \cite{ekomwenrenren2021hierarchical}.}
\label{fig:cpps01}
%\end{minipage}
\end{figure}

\begin{figure}[ht!]
\centering
\includegraphics[width=1\columnwidth]{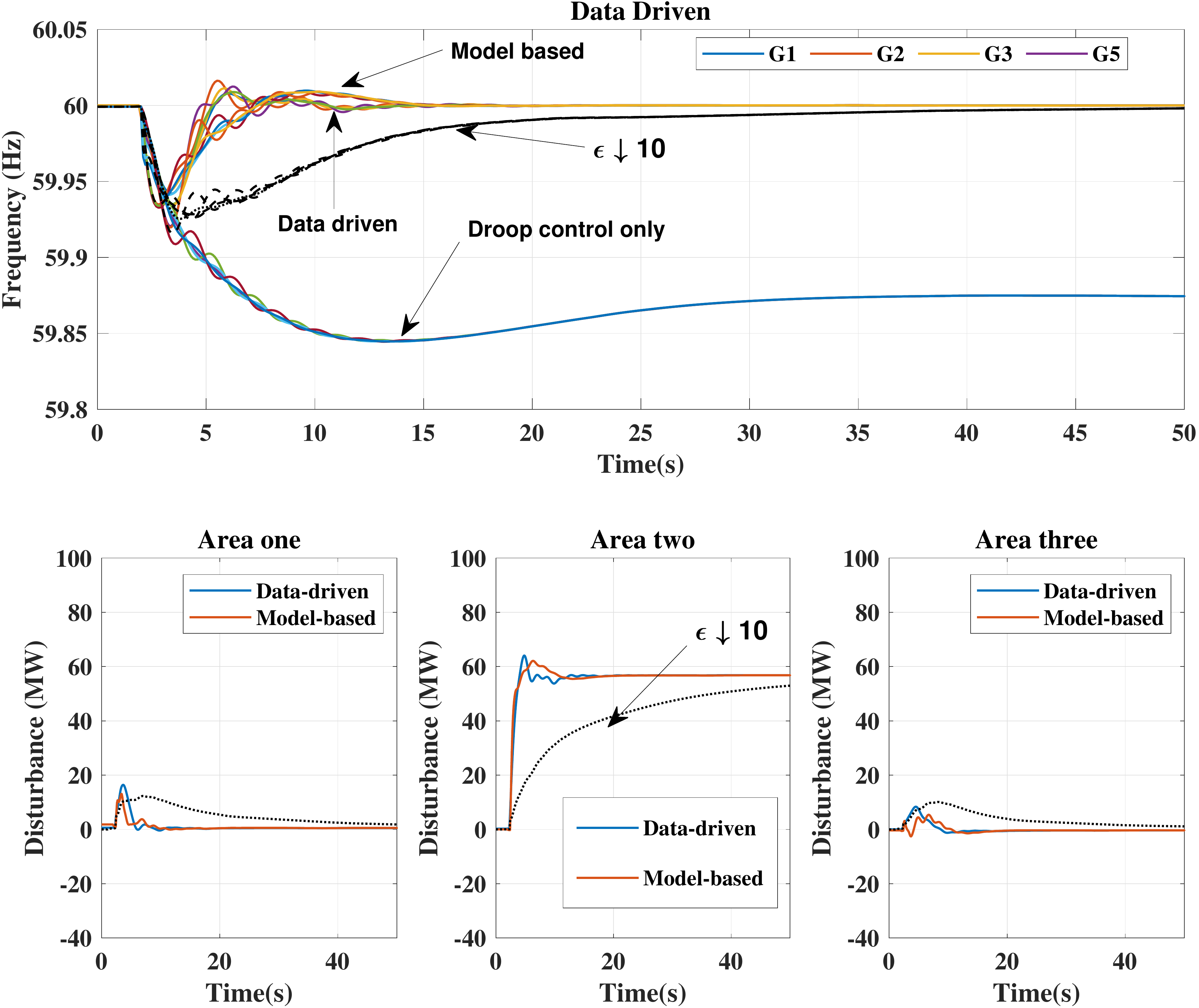}
\caption{Frequency and disturbance estimate during a 60 MW load change at bus 8 in Area 2.}
\label{fig:scenario1v1_freq}
\end{figure}

\begin{figure}[ht!]
\centering
\includegraphics[width=1\columnwidth]{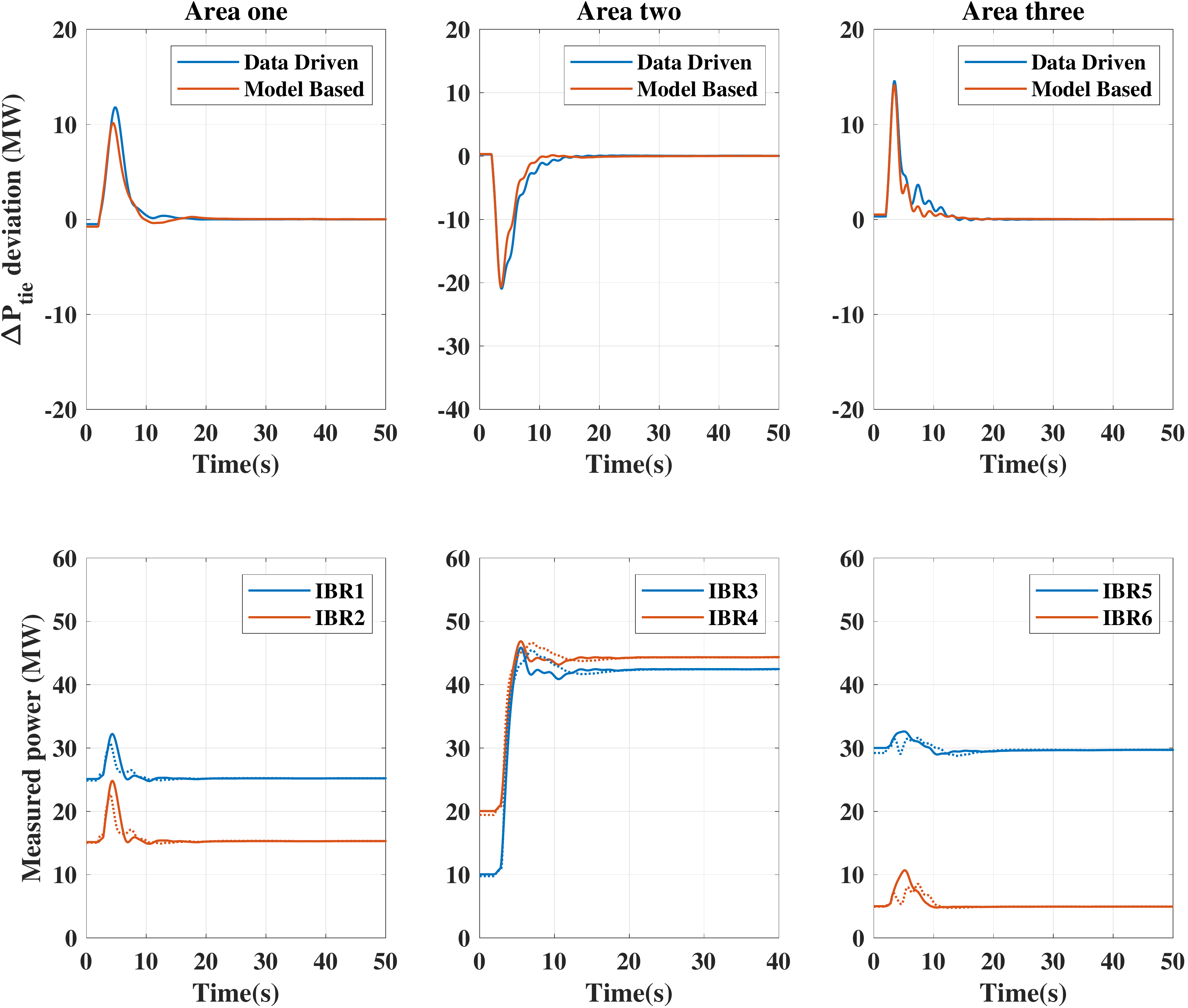}
\caption{Tie-line deviation and active power profiles during a 60 MW load change; dashed lines in the lower plots indicate the responses under model-based estimation.}
\label{fig:scenario1v1_pdev}
\end{figure}

\paragraph{Scenario \#1: Step Load Increase}
This scenario illustrates  the performance of the control scheme during step changes in loading. In the first case, a large load disturbance of 60 MW is introduced at $t = 2 s$ at bus 8 in Area 2. The frequency response and disturbance estimate of the system is shown in Figure \ref{fig:scenario1v1_freq}. Both the data-driven and model-based disturbance estimators localize the disturbance to Area 2, and quickly redispatch the IBRs in Area 2 (Figure \ref{fig:scenario1v1_pdev}) to compensate, with minimal transient response from the other areas. The frequency is quickly restored to the nominal value, with minimal oscillation. Similarly, the measured net tie-line deviation and IBR output power plots in Figure \ref{fig:scenario1v1_pdev} show a fast, non-oscillatory response. Due to the area-wise decentralized nature of the control scheme, the disturbance estimate, tie-line deviation and IBR outputs in the non-contingent areas return to their pre-disturbance values in steady-state. The black dotted lines overlaid on the responses in Figure \ref{fig:scenario1v1_freq} show the response when $\varepsilon$ is decreased by a factor of 10 to $\varepsilon = 0.01$. Tuning of the single scalar parameter $\varepsilon$ therefore allows for adjustment of the transient response.

%illustrate the effect of tuning the $\varepsilon$ parameter. We can modify the transient response by modifying $\varepsilon$, with the black dotted lines showing the slower response obtainable when $\varepsilon$ is reduced by a factor of 10 to $\varepsilon = 0.01$.

\begin{figure}[ht!]
\centering
\includegraphics[width=1\columnwidth]{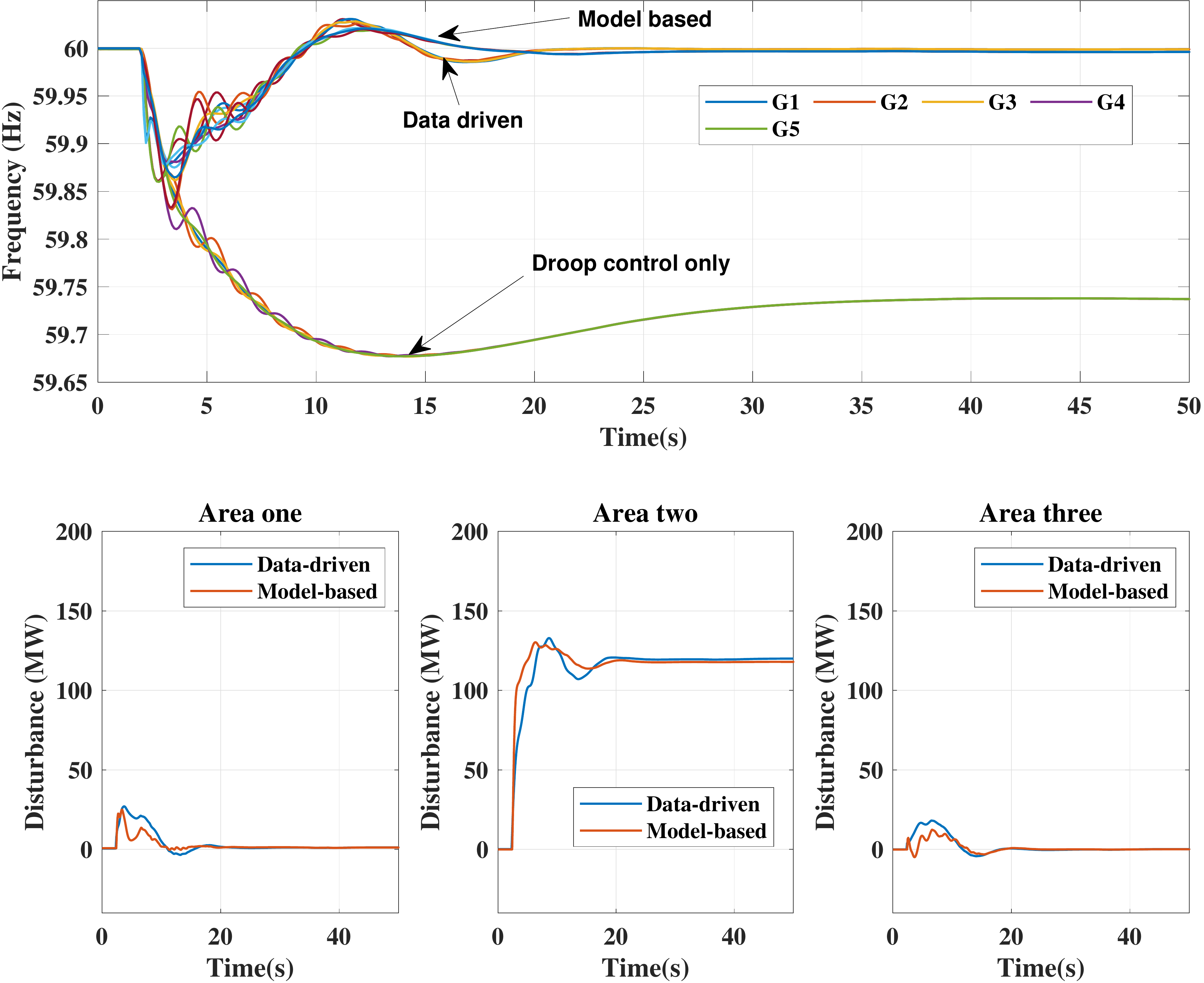}
\caption{Frequency and disturbance estimate during a 130 MW load change at bus 8 in Area 2.}
\label{fig:scenario1v2_freq}
\end{figure}

\begin{figure}[ht!]
\centering
\includegraphics[width=1\columnwidth]{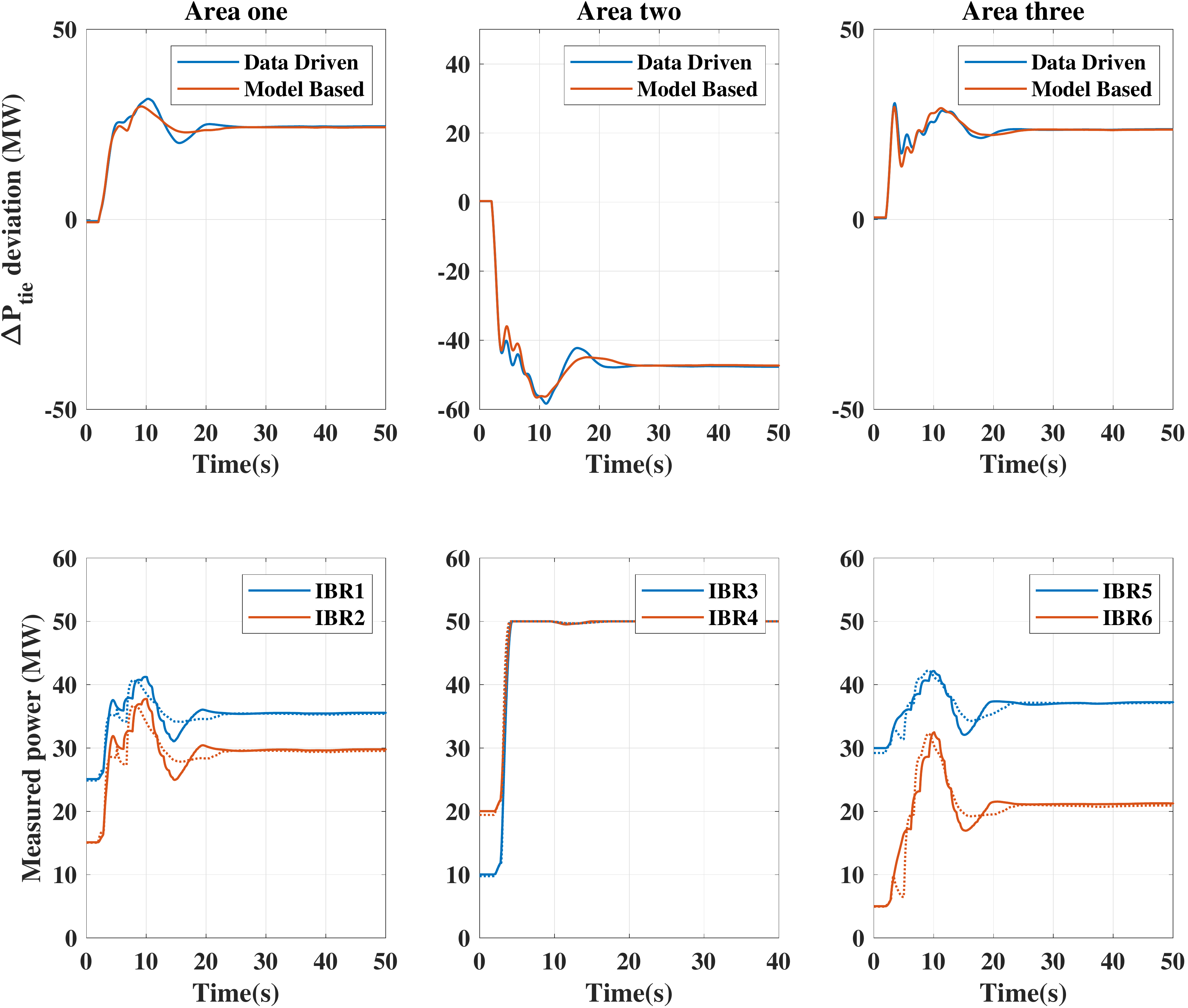}
\caption{Tie-line deviation and active power profiles during a 130 MW load change; dashed lines in the lower plots indicate the responses under model-based estimation.}
\label{fig:scenario1v2_pdev}
\end{figure}

Figures \ref{fig:scenario1v2_freq} and \ref{fig:scenario1v2_pdev} show the closed-loop response under a more severe load disturbance of 130 MW, again at bus 8 in Area 2. The disturbance exceeds the available spare IBR resources in the contingent area, and the inter-area coordination scheme proposed in \cite{ekomwenrenren2021hierarchical} is activated to facilitate support from IBRs in neighboring areas. While the details of these coordination scheme are not germane to our discussion here, the plots illustrate that the purely data-driven approach presented here produces similar results to a model-based approach. In summary, for both large and small load disturbances, our data-driven estimator is able to quickly spatially localize and compensate for a load disturbance, using absolutely no model information.

%Using both the data-driven and model-based disturbance estimator,the  system  frequency is regulated back  to its  nominal value. In both cases, the higher-level controller facilitates support from neighbors when the IBRs in the  contingent  area  reach  their  maximum  capacity. Simlar to the previous sufficient case, the data driven approach performs satisfactorily to the extreme load change in the system.

\paragraph{Scenario \#2: Heavy Renewable Penetration}

This scenario assesses the performance of the controller when one area of the power system is fed entirely by renewable energy sources. For this scenario, we have modified area three of the test system shown in Figure \ref{fig:cpps01}: in the modified system, generators G4 and G5 are replaced by a non-dispatchable IBRs with similar capacities and pre-disturbance dispatch points. A synchronous condenser is connected at bus 15 in area 3 to regulate the voltage by providing reactive power support. 

\begin{figure}[ht!]
\centering
\includegraphics[width=1\columnwidth]{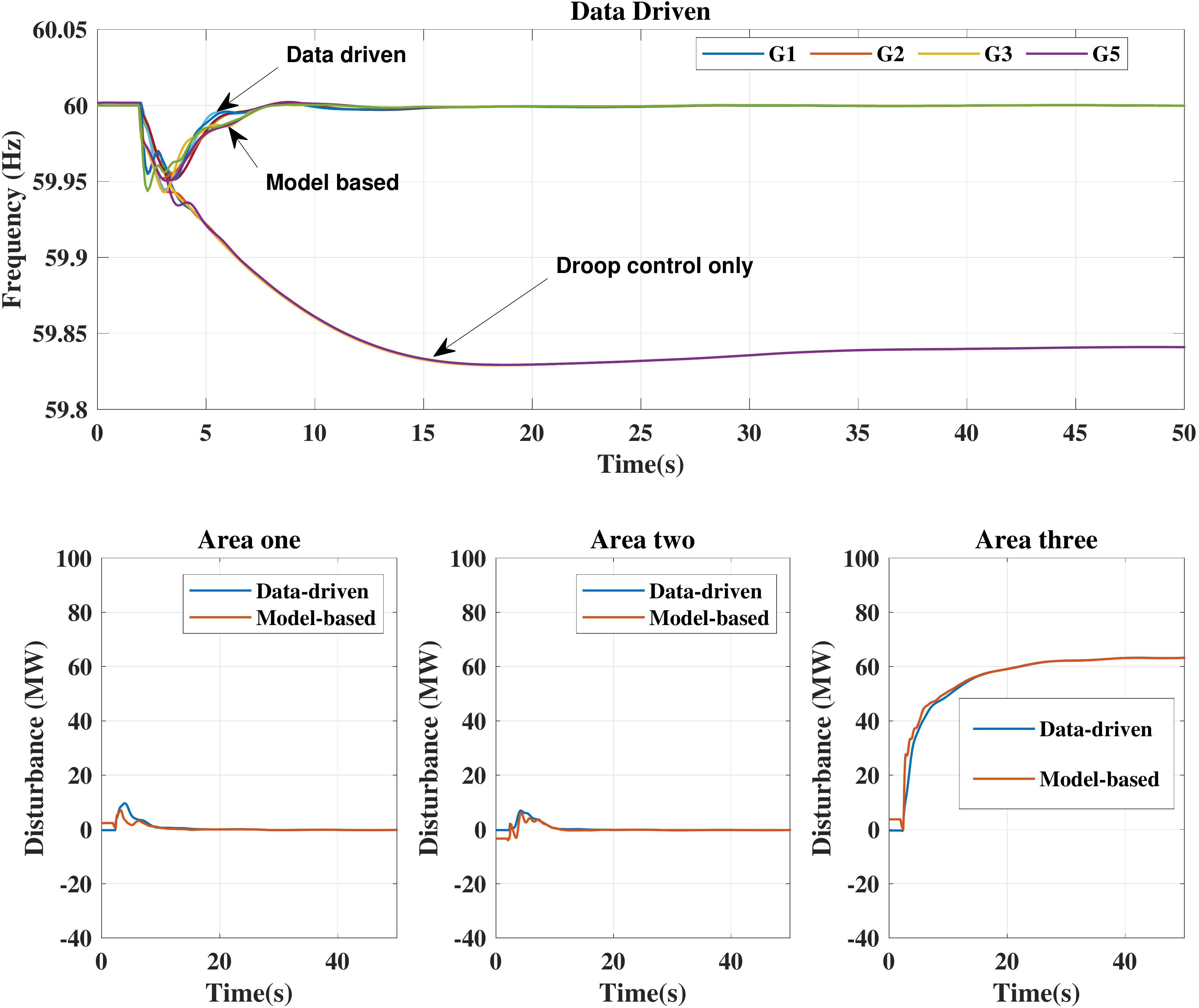}
\caption{Frequency response and disturbance estimates during combined step load change and a rate-limited change in wind speed.}
\label{fig:scenario2_freq}
\end{figure}

\begin{figure}[ht!]
\centering
\includegraphics[width=1\columnwidth]{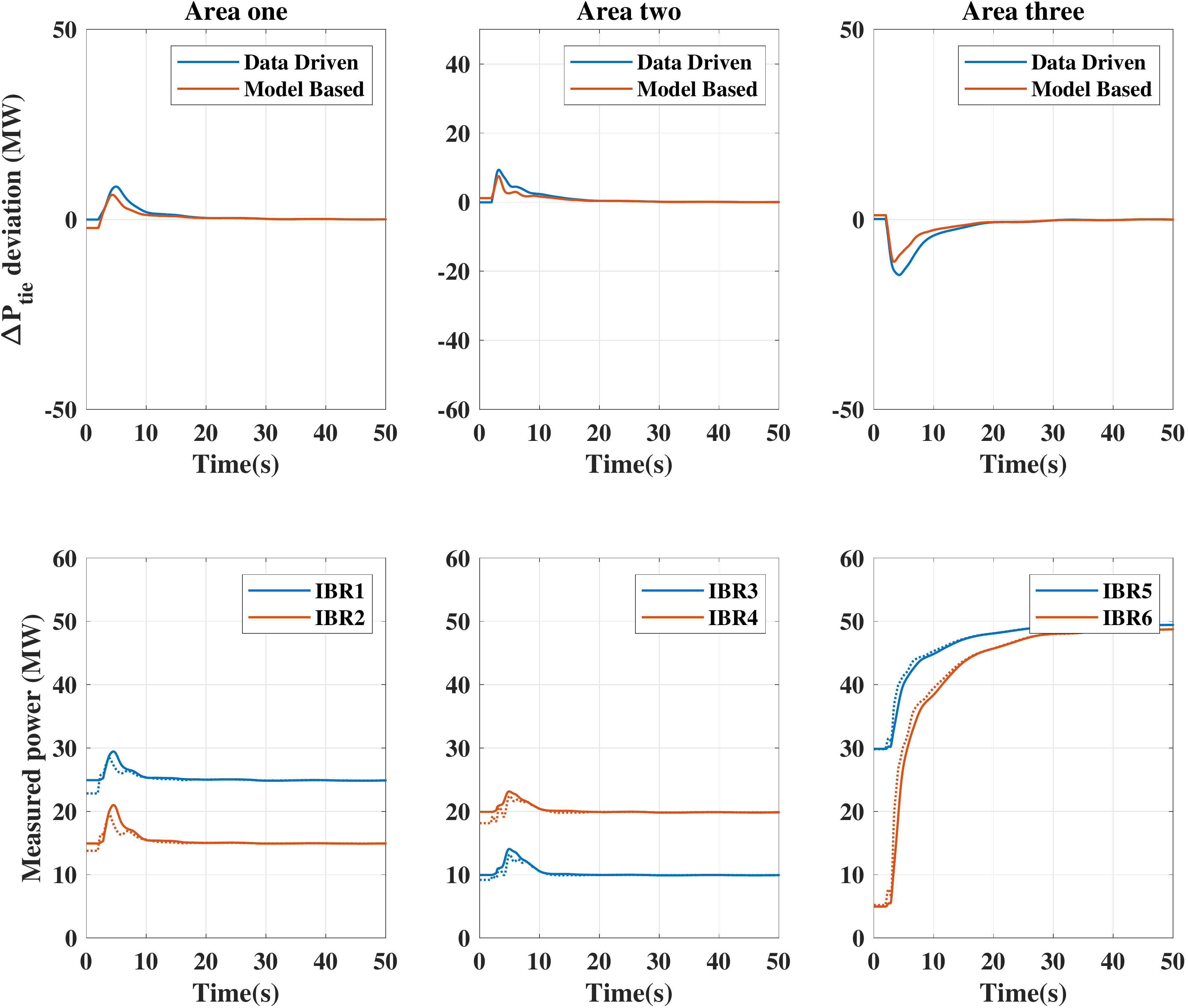}
\caption{Tie-line deviation and active power profiles during combined step load change and a rate-limited change in wind speed; dashed lines in the lower plots indicate the responses under model-based estimation.}
\label{fig:scenario2_pdev}
\end{figure}

For this scenario, a step disturbance of 40 MW is introduced at bus 14 in Area 3 at $t = 2$s. Simultaneously, a rate limited change in the wind speed to the wind turbine power system connected at bus 17 occurs. This is done to simulate the net-load variability and uncertainty in power supply introduced by renewables \cite{Ulbig2014}. From the responses shown in Figures \ref{fig:scenario2_freq}, \ref{fig:scenario2_pdev}, we observe that the IBRs in Area 3 respond quickly to the disturbance, and additionally, slowly ramp up their injections in response to the rate-limited decrease in the wind power. Throughout this process, the frequency is maintained very close to the nominal value. 

%\john{Etinosa, Figure \ref{fig:scenario2_pdev} looks like the wrong data is plotted; the IBRs in Area 2 are responding. I also think it would be clearer in this example if you stopped the wind speed adjustment sooner, so that the IBRs do not max out their capacities; I don't think we want to focus too much on the ``stage 2'' controller here.}

From this test, we conclude that the data-driven controller functions well in a low-inertia area, and even produces a closed-loop frequency response which is slightly improved compared to the crude model-based design from \cite{ekomwenrenren2021hierarchical}. %produces a slightly improved frequency response in terms of rise and settling times. Also, the disturbance estimate and IBR responses with the data-driven controller are smoother compared to the model-based. This is expected as the data-driven estimator captures more of the dynamics of the system compared to the simple two-state model used in designing the model-based estimator in \cite{ekomwenrenren2021hierarchical}.

\begin{figure}[ht!]
\centering
\includegraphics[width=1\columnwidth]{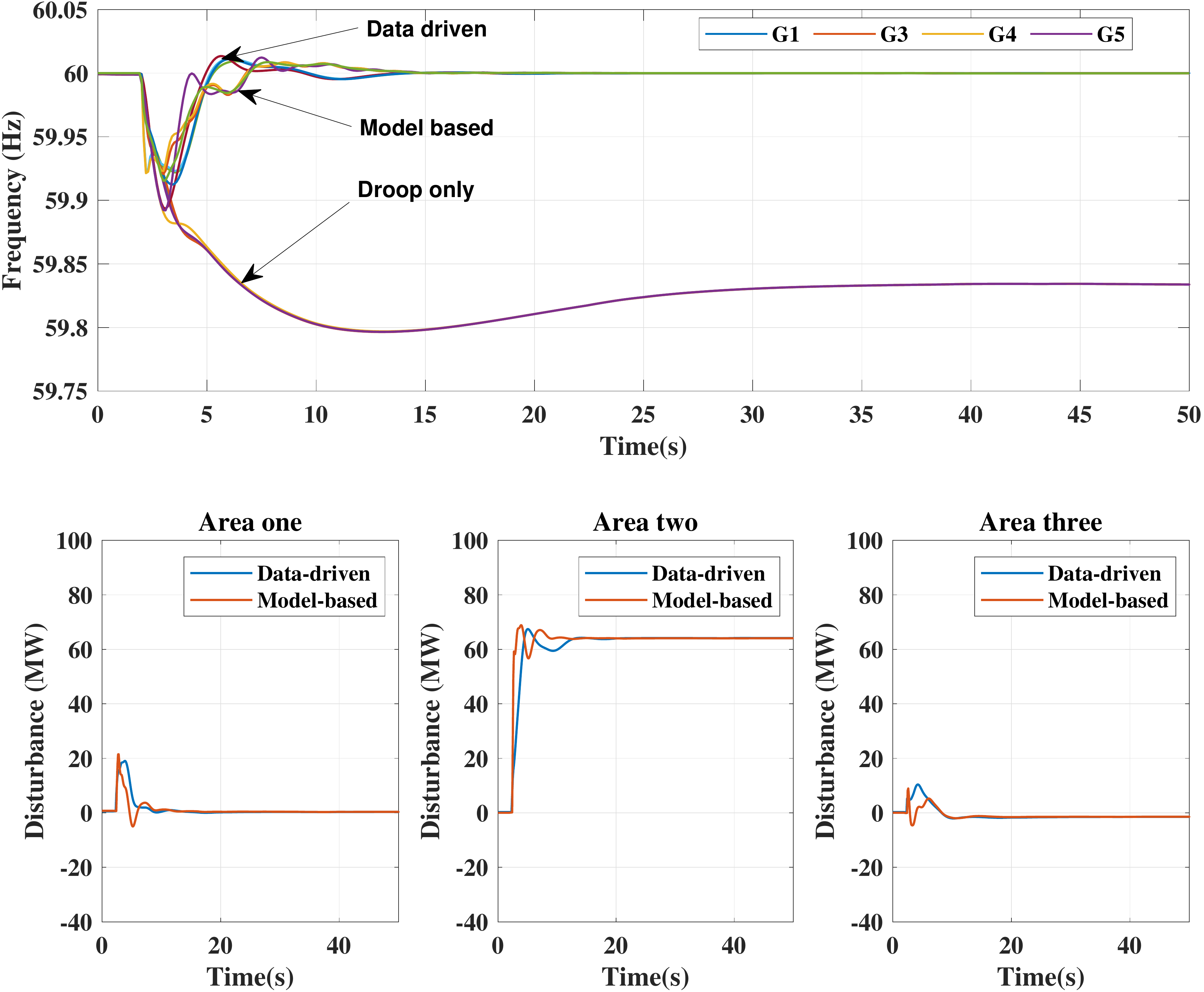}
\caption{Frequency and disturbance estimate during a loss of generator G2}
\label{fig:scenario3_freq}
\end{figure}

\paragraph{Scenario \#3: Generator Trip}
This scenario examines the performance of the controller during the loss of generator G2 in Area 2 at $t = 2$s. The response of the system to this loss is plotted in Figure \ref{fig:scenario3_freq}. Similar to Scenario \#2 above, we can observe that while the response of the controller under both data-driven and model-based disturbance estimation is quite fast and satisfactory, the data-driven controller also outperforms the model-based controller. This scenario illustrates the robustness of the method, as the collected data used for the design of the LCA controller was collected on the system \emph{including} the inertia and primary response of generator G2. Despite the model mismatch arising from the loss of the generator, the overall control response is similar to that in Scenario \#1. 

% \begin{figure}[ht!]
% \centering
% \includegraphics[width=1\columnwidth]{Images/DD_IREP_pdev_04.eps}
% \caption{Tie-line deviation and active power profiles during a loss of generator G2}
% \label{fig:scenario3_pdev}
% \end{figure}

\begin{figure}[ht!]
\centering
\includegraphics[width=1\columnwidth]{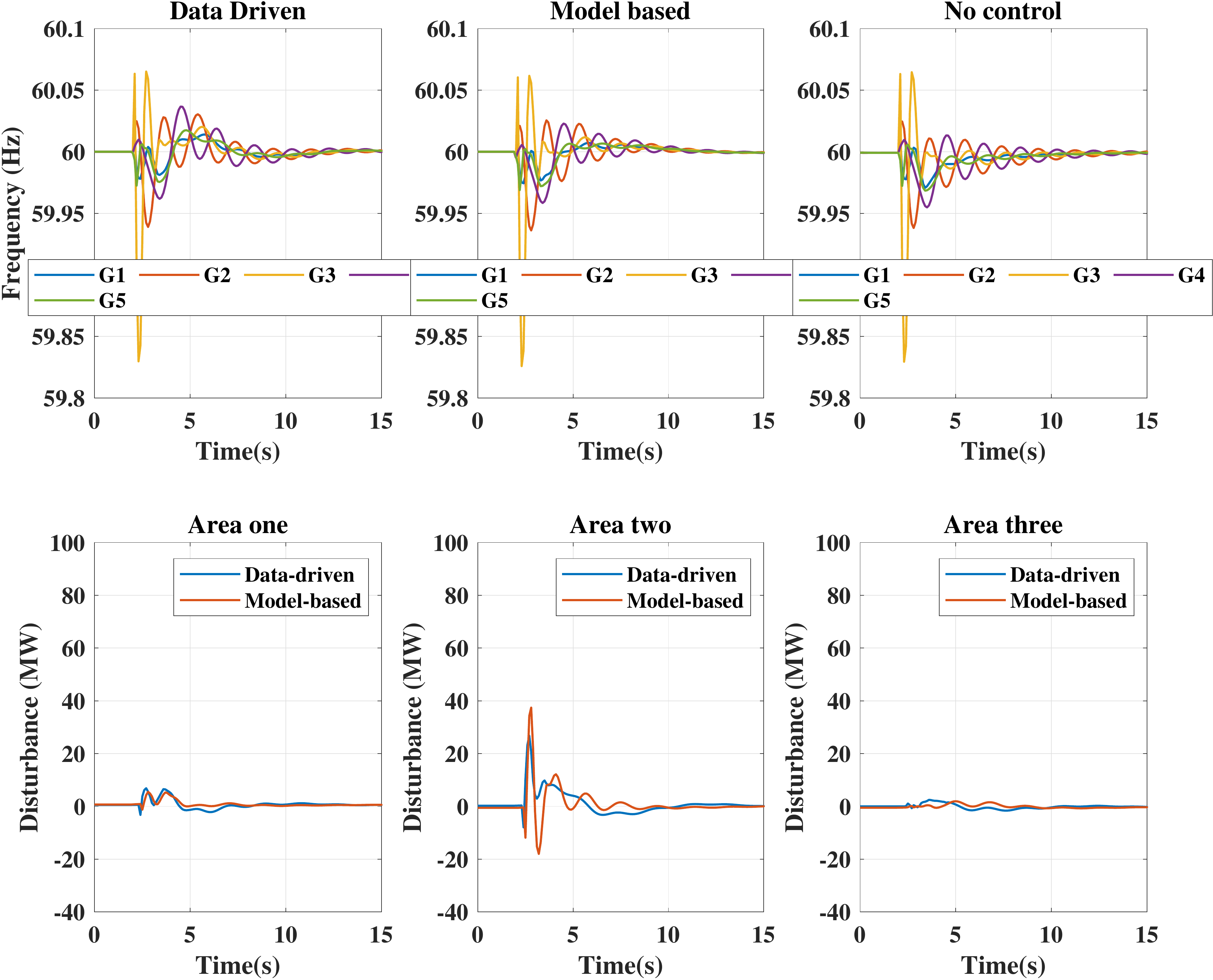}
\caption{Frequency and disturbance estimate during a three-phase fault in Area 2.}
\label{fig:scenario4_freq}
\end{figure}

\begin{figure}[ht!]
\centering
\includegraphics[width=1\columnwidth]{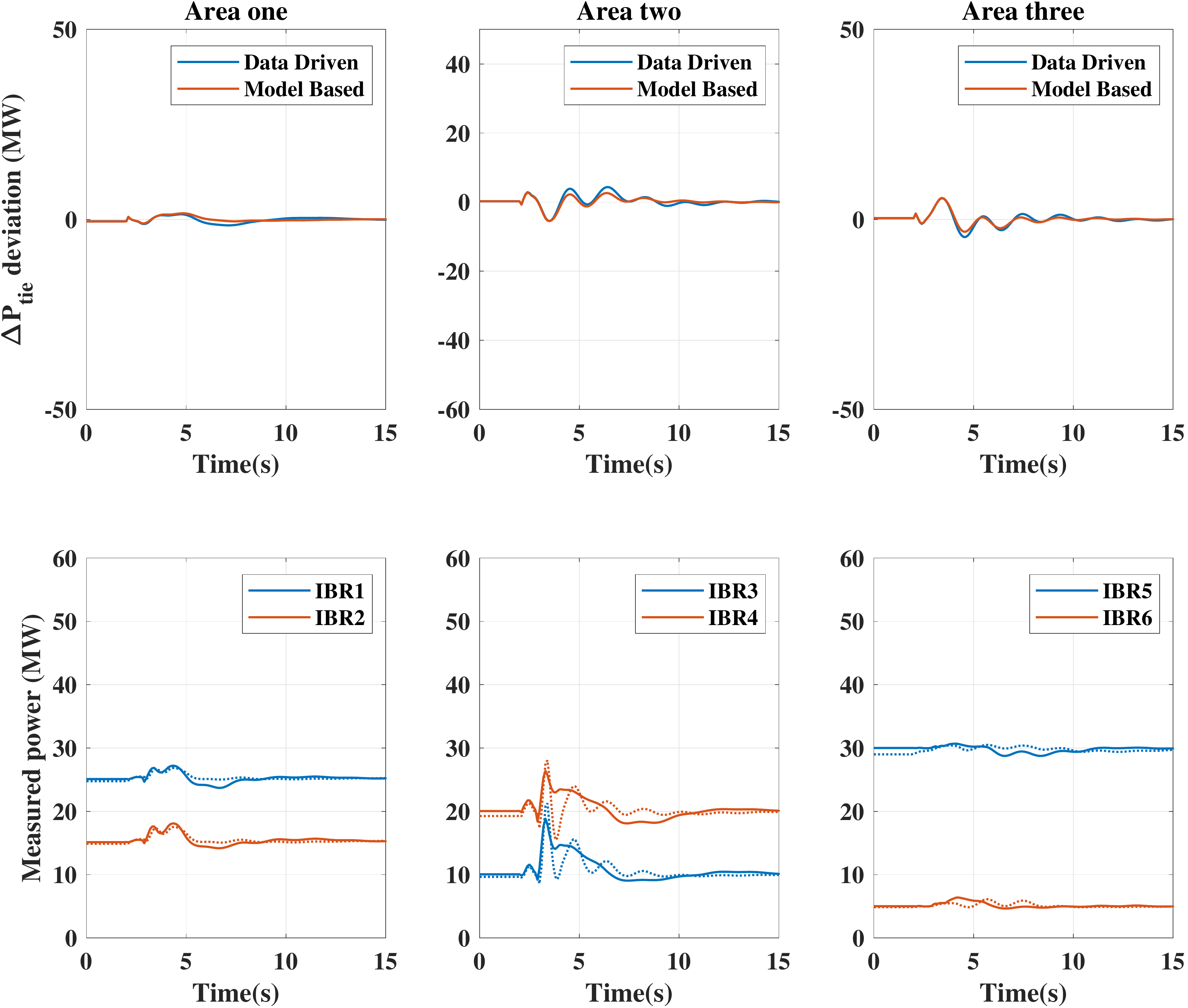}
\caption{Tie-line deviation and active power profiles during a three-phrase fault in Area 2; dashed lines in the lower plots indicate the responses under model-based estimation.}
\label{fig:scenario4_pdev}
\end{figure}

\paragraph{Scenario \#4: Three-Phase Fault}
The scenario assesses the performance of the controller during a symmetrical three-phase line-to-ground fault, which was introduced at bus 10 in Area 2 at $t = 2$s and cleared after $0.1$s. The response of the system is shown in Figures \ref{fig:scenario4_freq} and \ref{fig:scenario4_pdev}. It can be observed that the frequency response of the system with and without the controllers is very similar, with a small transient response in the disturbance estimate computed by the estimators. This indicates that the controller is able to effectively detect and respond to frequency events. The performance of the data-driven estimator in this scenario is satisfactory and is similar to the model-based estimator.

\begin{figure}[ht!]
\centering
\includegraphics[width=1\columnwidth]{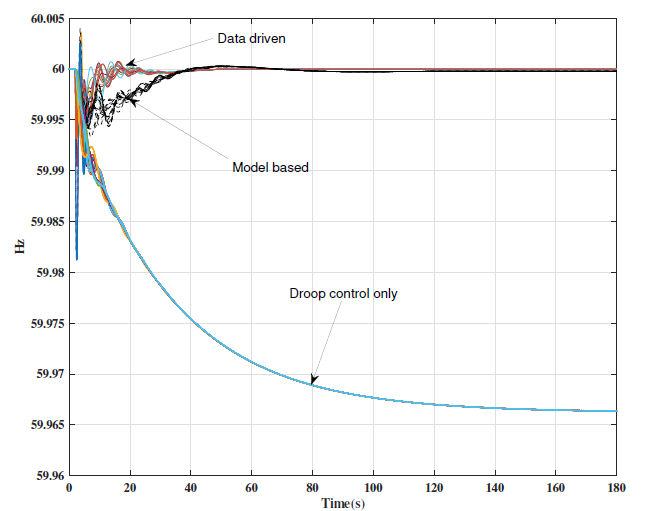}
\caption{Frequency response to a load change of 450 MW at bus 33.}
\label{fig:scenario5_freq}
\end{figure}

\begin{figure*}[ht!]
\centering
\includegraphics[width=1.8\columnwidth]{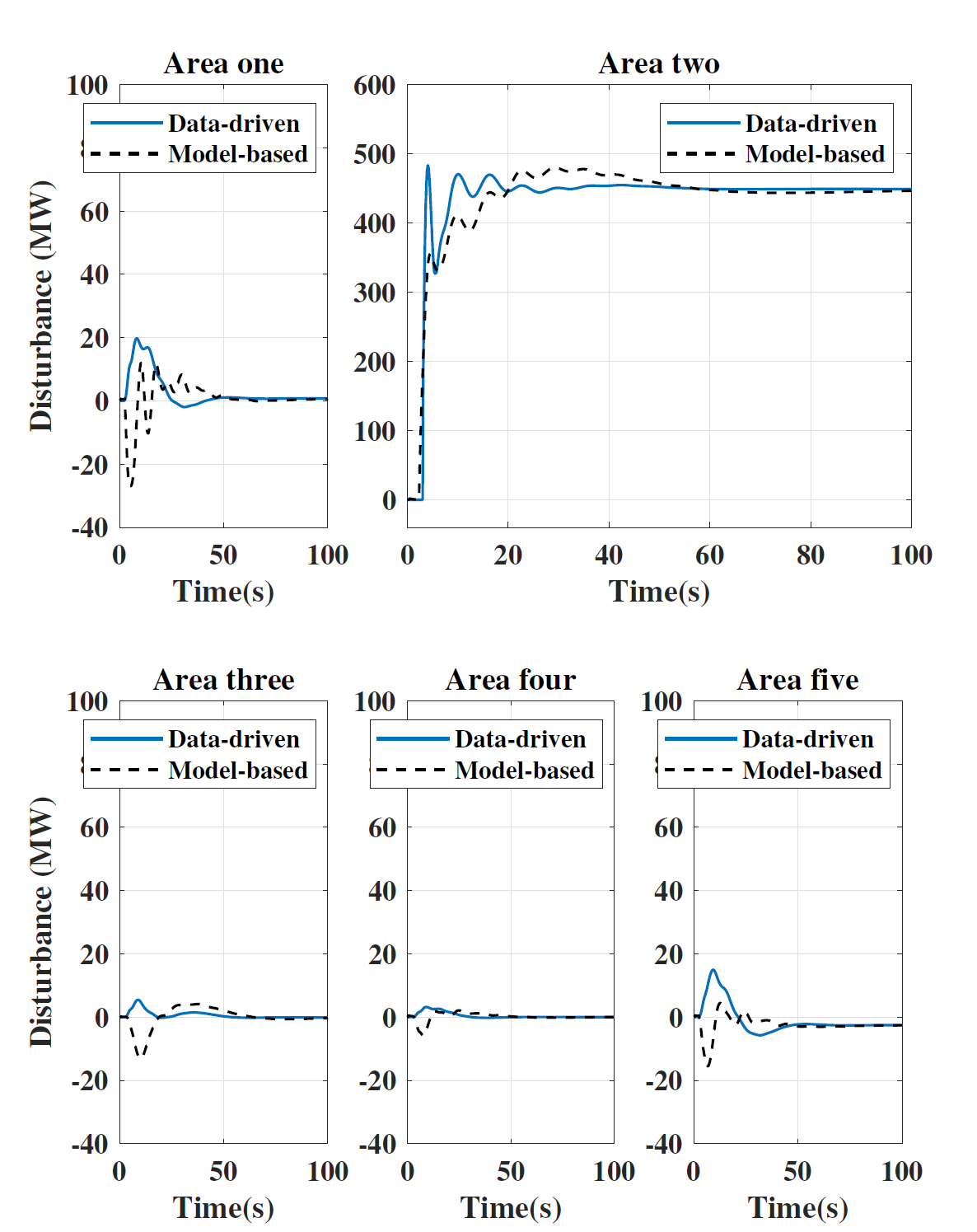}
\caption{Disturbance estimate during a load change of 450 MW at bus 33.}
\label{fig:scenario5_dis}
\end{figure*}

\paragraph{Scenario \#5: Step loading change in larger test system}
{This scenario is introduced to test the performance of the data-driven approach on the larger 5-LCA system shown in Figure \ref{fig:ieee68bus}, which is an IEEE benchmark model with 68 buses and is extensively described in \cite{pal2006robust}. 

Following the same data collection approach from Section \ref{Sec:TestSystem}, the perturbing set-point change for the IBRs in this system was as in \eqref{Eq:IBRPerturb}, with the noise term having a power spectral density of 1. For each LCA, measurements of frequency deviation and net inter-area power flow were collected for 10 seconds with a sampling period of 25 ms. The selections of $T_{\rm ini}$ and $\varepsilon$ were tuned for each area individually; in particular, it was found that larger values are beneficial for performance in larger areas. For instance, in the large NETS area the best tunings found were $T_{\rm ini} = 119$ and $\varepsilon = 0.3$, while in Area 4 the smaller values $T_{\rm ini} = 7$ and $\varepsilon = 0.01$ were found to be more appropriate. The intuition behind this is that larger areas are more dynamically complex, and require more data for initialization within our data-driven approach. 

\begin{figure}[ht!]
\centering
\includegraphics[width=1\columnwidth]{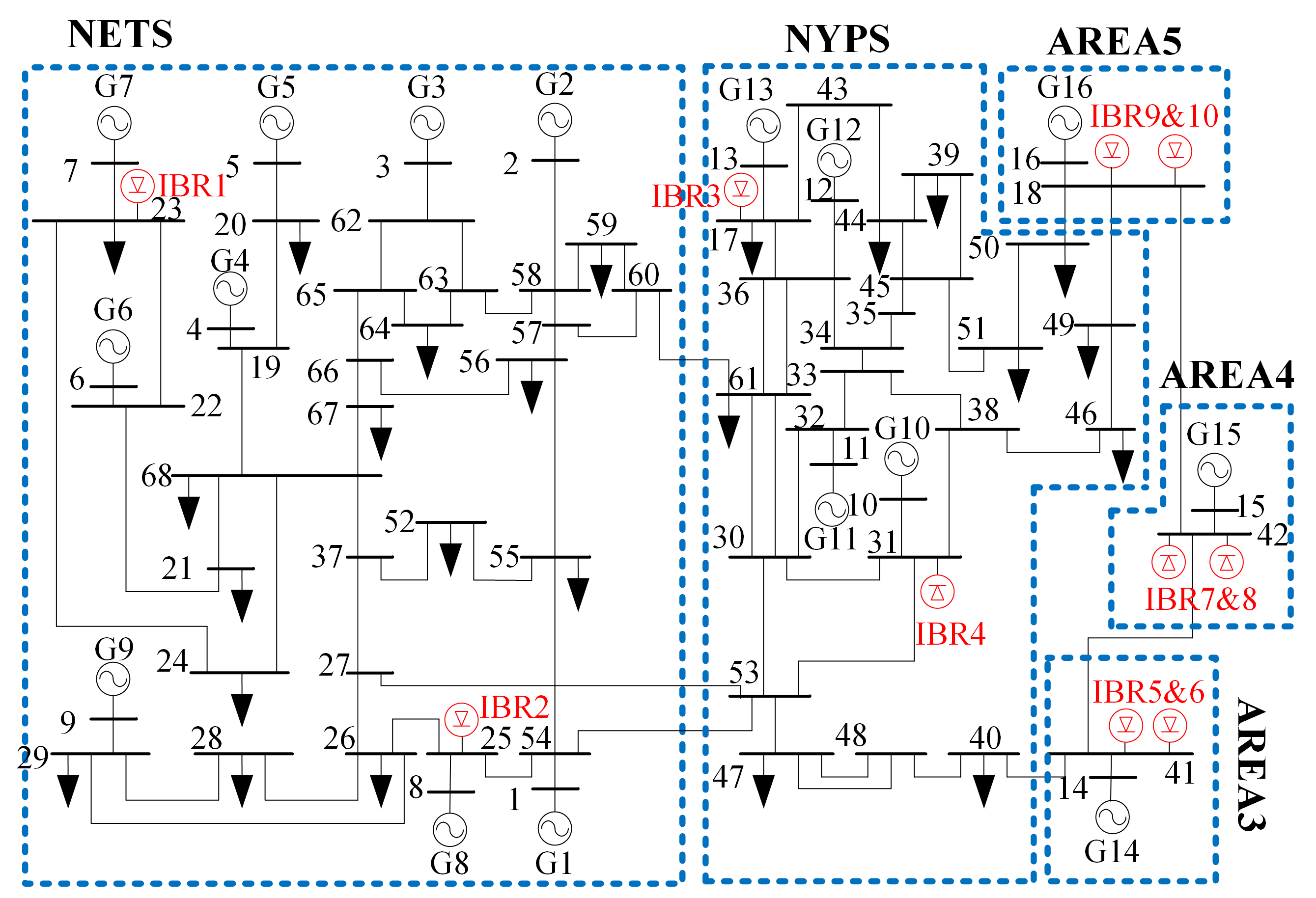}
\caption{ IEEE 68-bus test system \cite{ekomwenrenren2021hierarchical}.}
\label{fig:ieee68bus}
\end{figure}

For this case, we introduce a large step load change of 450 MW at bus 33 in the NYPS area at $t = 2$s. The frequency response of the system to this disturbance is plotted in Figure \ref{fig:scenario5_freq}, while Figure \ref{fig:scenario5_dis} shows the estimate of the load disturbance with the data-driven and model-based approaches. Similar to Scenarios \#2 and \#3 above, we can observe that while the response of the controller under both data-driven and model-based disturbance estimation is quite fast and satisfactory, the data-driven controller slightly outperforms the model-based controller. This is because the data-driven controller potentially captures more of the dynamics of the areas than the simplified model used for the model-based control approach.}

\section{Conclusion}
\label{Sec:Conclusion}
This paper has detailed the development of a data-driven area-based fast frequency control scheme, which extends the previous work in \cite{ekomwenrenren2021hierarchical} to be completely model-free. The controller rapidly redispatches inverter-based resources to compensate for local power imbalances within the bulk power system. The design is based only on recorded historical data, requires no model building or identification, and is tuned through adjustment of a single scalar parameter, making it potentially attractive for practical implementation. Simulation results on a nonlinear test system have been presented to validate the approach. Future work will be primarily concerned with 
minimizing the disruption to the system during the data collection phase via the design of improved excitation signals for IBRs, along with extension of the approach here to moving-horizon disturbance estimation. 


% % use section* for acknowledgment
% \section*{Acknowledgment}




%\bibliographystyle{IEEEtran}

\bibliographystyle{ieeetr}
\bibliography{brevaliasO,paper,comments,conference}


% \end{thebibliography}

\end{document}